\documentclass[12pt]{article}
\usepackage{graphicx}
\usepackage{natbib} 
\usepackage{url} 
\usepackage{amsmath} 
\usepackage{amsfonts}
\usepackage{amssymb} 
\usepackage{lineno}
\usepackage{setspace}
\usepackage{lscape} 
\usepackage{float} 
\usepackage{breakcites}
\usepackage{pdfpages}
\usepackage{xcolor}
\usepackage{hyperref}
\usepackage{setspace}

\graphicspath{{Figures/}}

\newcommand{\blind}{0}

\newcommand{\bs}[1]{\mathbf{#1}}

\begin{document}

\def\spacingset#1{\renewcommand{\baselinestretch}%
{#1}\small\normalsize} \spacingset{1}

\if0\blind
{
  \title{\bf PICAR: An Efficient Extendable Approach for Fitting Hierarchical Spatial Models}
  \author{Ben Seiyon Lee$^{1}$
    and \\
    Murali Haran$^{2}$\thanks{Corresponding Author: mharan@stat.psu.edu}\hspace{.2cm}\\
    $^{1}$Department of Statistics, George Mason University \\
    $^{2}$Department of Statistics, Pennsylvania State University \\
    }
  \maketitle
} \fi

\if1\blind
{
  \bigskip
  \bigskip
  \bigskip
  \begin{center}
    {\LARGE\bf PICAR: An Efficient Extendable Approach for Fitting Hierarchical Spatial Models}
\end{center}
  \medskip
} \fi

\begin{abstract}
Hierarchical spatial models are very flexible and popular for a vast array of applications in areas such as ecology, social science, public health,  and atmospheric science. It is common to carry out Bayesian inference for these models via  Markov chain Monte Carlo (MCMC). Each iteration of the MCMC algorithm is computationally expensive due to costly matrix operations. In addition, the MCMC algorithm needs to be run for more iterations because the strong cross-correlations among the spatial latent variables result in slow mixing Markov chains. To address these computational challenges, we propose a projection-based intrinsic conditional autoregression (PICAR) approach, which is a discretized and dimension-reduced representation of the underlying spatial random field using empirical basis functions on a triangular mesh. Our approach exhibits fast mixing as well as a considerable reduction in computational cost per iteration. PICAR is computationally efficient and scales well to high dimensions. It is also automated and easy to implement for a wide array of user-specified hierarchical spatial models. We show, via simulation studies, that our approach performs well in terms of parameter inference and prediction. We provide several examples to illustrate the applicability of our method, including (i) a high-dimensional cloud cover dataset that showcases its computational efficiency, (ii) a spatially varying coefficient model that demonstrates the ease of implementation of PICAR in the probabilistic programming languages {\tt stan} and {\tt nimble}, and (iii) a watershed survey example that illustrates how PICAR applies to models that are not amenable to efficient inference via existing methods. 
\end{abstract}

\noindent%
\vfill

\newpage
\spacingset{2} 

\section{Introduction}\label{Sec:Intro}
Hierarchical spatial models are commonly used to model spatial observations across many fields, for example species abundance in ecology, ice presence in glaciology, geo-referenced survey responses in public health studies, and crime incidence in urban areas. A quick search suggests that hierarchical spatial models are featured in thousands of research papers published annually. An important class of hierarchical spatial models is the spatial generalized linear mixed model (SGLMM). These are flexible models for both point referenced and areal data, where Gaussian random fields are used to model the spatial dependence across locations \citep{diggle1998model} . Other examples of hierarchical spatial models include spatially varying coefficient processes \citep{gelfand2003spatial,mu2018estimation}, covariate measurement error models \citep{xia1998spatio,bernadinelli1997disease,muff2015bayesian}, and co-regionalization models for multivariate responses \citep{banerjee2014hierarchical}. Hierarchical spatial models pose considerable computational challenges due to the large number of highly correlated spatial random effects, which can lead to costly likelihood evaluations and slow mixing in Markov Chain Monte Carlo (MCMC) algorithms. 

In this manuscript, we provide a computationally efficient approach for fitting high-dimensional hierarchical spatial models by reducing the correlation and the dimensions of the spatial random effects. What sets our projection-based intrinsic conditional autoregression (PICAR) approach apart from existing methods is: (i) our approach to reducing the correlation and dimensions of the random effects is efficient and automated; (ii) our approach is easily extendable, that is, it can be easily integrated into a hierarchical modeling scenario using implementations in the probabilistic programming languages {\tt stan} \citep{carpenter2017stan} and {\tt nimble} \citep{nimble2017}; and (iii) our method scales well to higher dimensional hierarchical spatial models than is typical of existing methods. A major advantage of PICAR is that in addition to providing an efficient estimation approach for large datasets, it is easy for non-experts to specify general hierarchical spatial models of their choice in this framework. 

Many innovative computational methods have been developed for high-dimensional spatial data in recent years \citep[cf.][]{cressie2008fixed,Banerjee2008Gaussian, higdon1998process, nychka2015multiresolution,Lindgren2011,katzfuss2017multi,datta2016hierarchical,Banerjee2013}. Recent studies \citep{heaton2019case, bradley2016comparison, sun2012geostatistics} examine several of these methods within the context of modeling high-dimensional spatial data. However, these methods primarily focus on linear spatial models with Gaussian observations. Notable exceptions are predictive process \citep{Banerjee2008Gaussian} and random projections \citep{Guan_Haran_2018}, which use a low-dimensional representation of the latent random field. However, these require updating the basis functions at each iteration of the MCMC algorithm. \citet{bradley2019bayesian} employs a basis representation of the latent random field while also exploiting conjugate distributions. However, the full conditional distributions may be difficult to construct as they require computing many matrices, vectors, and constants, and there are also open questions regarding the mixing of the resulting Gibbs samplers. Integrated Nested Laplace Approximations (INLA) \citep{rue2009approximate,lindgren2015bayesian} and Vecchia-Laplace approximations \citep{Zilber2021} provide very fast approximations of the posterior distribution. While these approaches are applicable to a wide array of models, they may not always be extendable to user-specified hierarchical spatial models. We provide more discussion about these algorithms later in the manuscript. 

Our method addresses the computational challenges by representing the spatial random effects with empirical basis functions. Various basis representations have been directly or indirectly used to model spatial data, for instance in the predictive process approach \citep{Banerjee2008Gaussian}, random projections \citep{Guan_Haran_2018, guan2019fast, Banerjee2013, park2019reduced}, Moran's basis for areal models \citep{hughes2013dimension}, stochastic partial differential equations \citep{Lindgren2011}, kernel convolutions \citep{higdon1998process}, eigenvector spatial filtering \citep{griffith2003spatial}, multi-resolution radial basis functions \citep{nychka2015multiresolution,katzfuss2017multi}, and principal components \citep{higdon2008computer,cressie2015statistics}, among others. We utilize a set of basis functions inspired by the Moran's I statistic and piece-wise linear basis functions. To our knowledge, this is the first approach that readily lends itself to user-specified hierarchical spatial models while also remaining computationally efficient for large datasets. We demonstrate the applicability of PICAR via simulation studies as well as applications to two high-dimensional environmental datasets. 

The rest of the paper is organized as follows. In Section \ref{Sec:SpatialModels}, we describe hierarchical spatial models and discuss their computational challenges. In Section \ref{Sec:OurApproach}, we describe our PICAR approach in detail. In Section \ref{Sec:SimulationStudy}, we present simulated examples for: (i) a spatial model for binary observations; (ii) a spatially varying coefficient model for count observations, implemented in PICAR using \texttt{stan} and \texttt{nimble}; and (iii) a model for ordered categorical spatial data that cannot be fit using existing publicly available code but can be easily fit using PICAR. In Section \ref{Sec:Applications}, we apply PICAR to two large spatial datasets: cloud cover from satellite imagery and water quality ratings in Maryland watersheds. Finally, we provide a summary and directions for future research in Section \ref{Sec:Discussion}.
\section{Hierarchical Spatial
Models}\label{Sec:SpatialModels}

We begin by describing a general framework for hierarchical spatial models and provide several examples that will be explored in depth via simulation studies. We also provide a general discussion of the computational challenges for fitting these models. 

\subsection{Model Specification}\label{SubSec:SGLMM}
Let $Z(\bs{s})$ denote a spatial process at location $\bs{s}$ in a spatial domain $\mathcal{D}\subset \mathbb{R}^{2}$ where $d$ is typically 2 or 3. We define $Z(\bs{s})$ as:
\begin{equation}\label{EQ:SLMM}
  Z(\bs{s})=\mathbf{X}'(\bs{s})\pmb{\beta} + w(\bs{s}) + \epsilon(\bs{s}), \mbox{ for }\bs{s}\in \mathcal{D},
\end{equation}
where $\mathbf{X}(\bs{s})$ is a $k$-dimensional vector of covariates associated with location $\bs{s}$ and $\pmb{\beta}$ is a $k$-dimensional vector of coefficients. The micro-scale measurement errors or nugget are modeled as an uncorrelated Gaussian process with zero mean and variance $\tau^{2}$ where $\epsilon(\bs{s})\sim \mathcal{N}(0,\tau^{2})$ for $\bs{s}\in\mathcal{D}$.

We impose spatial dependence in the observations $\mathbf{Z}=\{Z(\bs{s}): \bs{s}\in \mathcal{D}\}$ by modeling the spatial random effects $\mathbf{W}=\{w(\bs{s}): \bs{s}\in \mathcal{D}\}$ as a stationary zero-mean Gaussian process with a positive definite covariance function $\bs{C}(\pmb{\Theta})$ with covariance function parameters $\pmb{\Theta}$. For a finite set of locations $\{\bs{s}_{1},...,\bs{s}_{n}\}$, the spatial random effects $\mathbf{W}$ are distributed as a multivariate normal distribution $\mathbf{W}|\pmb{\Theta} \sim \mathcal{N}(\bs{0},\bs{C}(\pmb{\Theta}))$ and the covariance matrix $\bs{C}(\pmb{\Theta})_{ij}=\mbox{Cov}(w(\bs{s}_{i}),w(\bs{s}_{j}))$. The Mat\'ern covariance function is a widely used class of stationary and isotropic covariance functions \citep{stein2012interpolation} with parameters $\pmb{\Theta}=(\sigma^{2},\phi,\nu)$ such that:
\begin{equation}\label{EQ:Matern}
    \bs{C}(\bs{s}_{i},\bs{s}_{j};\pmb{\Theta})=\sigma^{2}\frac{1}{\Gamma(\nu)2^{\nu-1}}\Big(\sqrt{2\nu}\frac{h}{\phi}\Big)^{\nu}K_{\nu}\Big(\sqrt{2\nu}\frac{h}{\phi}\Big),
\end{equation}
where $h=||\bs{s}_{i}-\bs{s}_{j}||$ is the Euclidean distance between locations $\bs{s}_{i}$ and $\bs{s}_{j}$, $\sigma^{2}>0$ is the partial sill or scale parameter of the process, $\phi>0$ is the range parameter for spatial dependence, and $\nu$ is the smoothness parameter, which is commonly fixed prior to model-fitting. $K_{\nu}(\cdot)$ is the modified Bessel function of the second kind. 

Hierarchical spatial models may be broadly described as   \citep[cf.][]{wikle1998hierarchical}:
\begin{align*}
    \mbox{\textbf{Data Model:}}&\qquad \mbox{observations } \textbf{Z}| \mathbf{W}, \mbox{ data model parameters}\\
    \mbox{\textbf{Process  Model:}}&\qquad\mathbf{W} | \mbox{ covariance function parameters }\\
    \mbox{\textbf{Parameter  Model:}}&\qquad  \mbox{priors for parameters}
\end{align*}

In Sections \ref{Sec:SimulationStudy} and \ref{Sec:Applications} we discuss the implementation of PICAR in the context of several examples of hierarchical spatial models, including spatial generalized linear mixed models for non-Gaussian data, a spatially varying coefficients model for count data, and a cumulative-logit model for ordered categorical data. 

\subsection{Model Fitting and Computational Challenges}\label{Sec:Challenges}
The computational challenges are rooted in the dimensionality and correlation of the spatial random effects $\bf{W}$. Hierarchical spatial models typically require a costly evaluation of an $n-$dimensional multivariate normal likelihood function ($\mathcal{O}(n^{3})$) at each iteration of the MCMC algorithm. Moreover, highly correlated spatial random effects can lead to poor mixing in MCMC algorithms \citep[cf.][]{Christensen2006robust,haran2003accelerating}. 

There is a large literature on addressing computational challenges in spatial models though the vast majority of methods are focused on linear Gaussian spatial models. 
Popular approaches include low-rank approximations \citep{cressie2008fixed,Banerjee2013}, compact support or covariance tapering \citep{furrer2006covariance, stein2013statistical}, multiresolution approaches \citep{nychka2015multiresolution,katzfuss2017multi}, sparse representations of the $n \times n$ precision matrix via spatial partial differential equations \citep{Lindgren2011}, and nearest-neighbor Gaussian processes \citep{datta2016hierarchical}. These typically focus on the marginal distribution of the spatial observations $\mathbf{Z}=\{Z(\bs{s}): \bs{s}\in \mathcal{D}\}$, where  $\mathbf{Z}|\pmb{\beta},\pmb{\Theta} \sim \mathcal{N}(\bs{X}\pmb{\beta},\bs{C}(\pmb{\Theta}))$ after integrating out the latent spatial random effects $\mathbf{W}$. Hence, they are not easily extended to more complex hierarchical spatial models. 

For hierarchical spatial models with non-Gaussian observations (e.g. SGLMMs), \citet{sengupta2013hierarchical} and \citet{sengupta2016predictive} extend fixed-rank kriging \citep{cressie2008fixed} to non-Gaussian satellite imagery by: (1) representing the spatial random effects using bi-square basis functions; (2) estimating the model parameters via the expectation-maximation (EM) algorithm; and (3) embedding Laplace approximations in the E-step to improve computational efficiency. The predictive process approach \citep{Banerjee2008Gaussian} also implements a basis representation of the spatial random effects $\mathbf{W}\in \mathbb{R}^{n}$. Prior to model fitting, $m$ reference locations or knots $\mathbf{s^{*}}=\{\bs{s}_{1}^{*},\bs{s}_{2}^{*},...,\bs{s}_{m}^{*}\}$ are selected where $m\ll n$. Predictive processes approximates the spatial random effects such that $\mathbf{W}\approx \bs{C}(\mathbf{s},\mathbf{s}^{*})\bs{C}^{*-1}\mathbf{W}^{*}$, where $\mathbf{W}^{*}$ is an m-dimensional vector of the realizations of the Gaussian random field at knot locations $\mathbf{s^{*}}$, $\bs{C}^{*}$ represents the $m\times m$ covariance matrix corresponding to the knot locations, and $\bs{C}(\mathbf{s},\mathbf{s}^{*})$ denotes the cross-covariance between observation locations $\mathbf{s}$ and the knot locations $\mathbf{s}^{*}$. Knots can be selected using an adaptive approach based on point processes \citep{guhaniyogi2011adaptive}. Note that the computational speedup comes from estimating a lower-dimensional set of spatial random effects $\mathbf{W}^{*}$ and an $m \times m$ covariance matrix $\bs{C}^{*}$. However, the basis functions matrix $\bs{C}(\mathbf{s},\mathbf{s}^{*})\bs{C}^{*-1}$ must be constructed at each iteration of the algorithm, which incurs a cost of roughly $O(m^3 + nm^2)$. In PICAR, the basis functions are constructed just once, and the costs for each iteration are linear in $p$, the dimension of the basis coefficients.

\citet{Guan_Haran_2018} use random projections to generate approximate eigenvector basis functions. The basis functions approximate the true eigencomponents of the Mat\'ern class of covariance functions. The dominant computational cost is driven by large matrix-to-matrix multiplications, which can be easily parallelized. The spatial random effects are approximated as $\mathbf{W}\approx \tilde{\bs{U}}_{m}\tilde{\bs{D}}_{m}^{1/2}\pmb{\delta}$, where $\tilde{\bs{U}}_{m}$ and $\tilde{\bs{D}}_{m}$ are the first $m$ approximate eigencomponents of the covariance matrix $\bs{C}(\pmb{\Theta})$, and $\pmb{\delta}$ is the $m$-dimensional vector of reparameterized spatial random effects. While this approach bypasses knot selection, it still requires repeated constructions of the approximate eigenvector basis functions with a cost of $O(m^{3}+n
m^{2})$.

Reparameterization approaches \citep{Christensen2006robust, haran2003accelerating, Guan_Haran_2018} can reduce correlation in the spatial random effects, which often results in faster mixing MCMC algorithms. However, these techniques can be very expensive for high-dimensional data, since the reparameterization step is expensive and the number of spatial random effects remain unchanged. Data augmentation approaches \citep{DeOliveira2000, Albert1993} apply to some classes of hierarchical models, resulting in a Gibbs sampler for the spatial random effects, but this still requires large matrix operations on dense covariance matrices, and does not necessarily address mixing issues in the resulting MCMC algorithm.

\section{PICAR Approach}\label{Sec:OurApproach}
In this section, we present our projection-based intrinsic conditional autoregression (PICAR) approach that is designed to efficiently fit hierarchical spatial models. In this framework, we represent the spatial random effects $\mathbf{W}=(w(\bs{s}_{1}),w(\bs{s}_{2}),...,w(\bs{s}_{n}))'$ as a linear combination of basis functions:
$$\mathbf{W}\approx \mathbf{\Phi}\pmb{\delta}, \qquad \pmb{\delta}\sim \mathcal{N}(\bs{0},\pmb{\Sigma}_{\pmb{\delta}}),$$
 where $\mathbf{\Phi}$ is an $n\times p$ basis function matrix where each column contains a spatial basis function, $\pmb{\delta}\in\mathbb{R}^p$ are the re-parameterized spatial random effects (or basis coefficients), and $\pmb{\Sigma}_{\pmb{\delta}}$ is the $p\times p$ covariance matrix for the coefficients. Basis functions can be interpreted as a set of distinct spatial patterns that can be used to construct a spatial random field, along with their coefficients. Basis representation has been a popular approach to model spatial data \citep[cf.][]{cressie2008fixed,Banerjee2008Gaussian,hughes2013dimension,Lindgren2011,rue2009approximate,Christensen2006robust,haran2003accelerating,griffith2003spatial,higdon1998process,nychka2015multiresolution,mak2018efficient, higdon2008computer}. Examples of basis functions include splines, wavelets, empirical orthogonal functions, combinations of sines and cosines, piece-wise linear functions, and many others. Basis representations tend to be computationally efficient as they help bypass large matrix operations, reduce the dimensions of the spatial random effects $\mathbf{W}$, and as in our case, reduces correlation in $\mathbf{W}$.
 
 Basis representation approaches are subject to key limitations. Radial, Fourier, wavelets, and Gaussian kernels can be overcomplete, meaning that the number of basis functions can be much larger than the number of observations; thereby increasing computational costs. Radial basis functions, Gaussian kernels, and predictive processes rely on careful placement of knot locations as well as specifying the bandwidth of the basis functions. Predictive processes or random projections still require large matrix operations, though not at the scale of the full hierarchical model. Empirical orthogonal functions, or principal components, require replicate observations at the same set of locations, which may not be readily available. The eigenvectors of a Mat\'ern covariance function (see Section 4.2) provide a robust set of spatial basis functions, but these require many pre-specified tuning parameters such as the spatial range $\phi$, partial sill $\sigma^{2}$, smoothness $\nu$, and rank.
 
\subsection{Projection-based Intrinsic Conditional Auto-Regression (PICAR)}\label{SubSec:PICAR}

The PICAR approach can be outlined as follows:
\begin{enumerate}
    \item Generate a triangular mesh on the spatial domain  $\mathcal{D}\subset \mathbb{R}^{2}$.
    \item Construct a spatial field on the mesh nodes using data-driven basis functions.
    \item Interpolate onto the observation locations using piece-wise linear basis functions.
\end{enumerate}
We provide details for each step below.
\subsubsection*{Mesh Construction}
Prior to fitting the model, we generate a mesh enveloping the observed spatial locations via Delaunay Triangulation \citep{hjelle2006triangulations}. Here, we divide the spatial domain $\mathcal{D}$ into a collection of non-intersecting irregular triangles. The triangles can share a common edge, corner (i.e. nodes or vertices), or both. The mesh generates a latent undirected graph $\bs{G}=\{\bs{V},E\}$, where $\bs{V} =\{\bs{v}_1,\bs{v}_2, . . . , \bs{v}_m\}$ are the locations of the mesh vertices and $E$ are the edges. Each edge $E$ is represented as a pair $(i, j)$ denoting the connection between $\bs{v}_i$ and $\bs{v}_j$. The graph $\bs{G}$ is characterized by its adjacency or neighborhood matrix $\bs{N}$, an $m\times m$ matrix where $\bs{N}_{ii}=0$, $\bs{N}_{ij} =1$ when mesh node $\bs{v}_i$ is connected to node $\bs{v}_j$ and $\bs{N}_{ij} =0$ otherwise for $\bs{v}_i \neq \bs{v}_j$. The triangular mesh is built using the \textbf{R-INLA} package \citep{lindgren2015bayesian}. Guidelines for mesh construction are provided in \citet{lindgren2015bayesian}, and details pertaining to algorithms for Delaunay triangulation can be found in \citet{hjelle2006triangulations}.  

For the PICAR approach, we suggest practitioners model the data using varying mesh densities such as $m=\lfloor\gamma n\rfloor$ for $\gamma=\{0.1,0.2,...,1.9,2\}$ and sample size $n$, and then choose $m$ based on the lowest out-of-sample prediction error. Please see the Supplement (Section 5) for a numerical study that examines how mesh density affects prediction.

\subsubsection*{Moran's Basis Functions}\label{SubSec:Moran}
We generate a spatial random field on the set of mesh vertices $\bs{V}$ of graph $\bs{G}$ using the Moran's basis functions \citep{hughes2013dimension, griffith2003spatial}. \citet{griffith2003spatial} proposes an augmented hierarchical spatial model (spatial eigenfiltering) using a subset of eigenvectors of the Moran's operator $\mathbf{(I-11'/}m\mathbf{)N(I-11'/}m)$, where $\mathbf{N}$ is the $m \times m$ neighborhood matrix, $\mathbf{I}$ is the $m$-dimensional identity matrix and $\mathbf{1}$ is an $m$-dimensional vector of $1$'s. The operator appears in the numerator of the Moran's I statistic, which is a diagnostic for spatial dependence \citep{moran1950notes} typically used for areal spatial data. Positive eigencomponents of the Moran's operator correspond to varying magnitudes and patterns of positive spatial dependence. For the triangular mesh, the positive eigenvectors represent the patterns of spatial dependence among the mesh nodes, and their corresponding eigenvalues denote the magnitude of spatial dependence.

\begin{figure}[ht]
    \centering
    \includegraphics[width=0.75\textwidth]{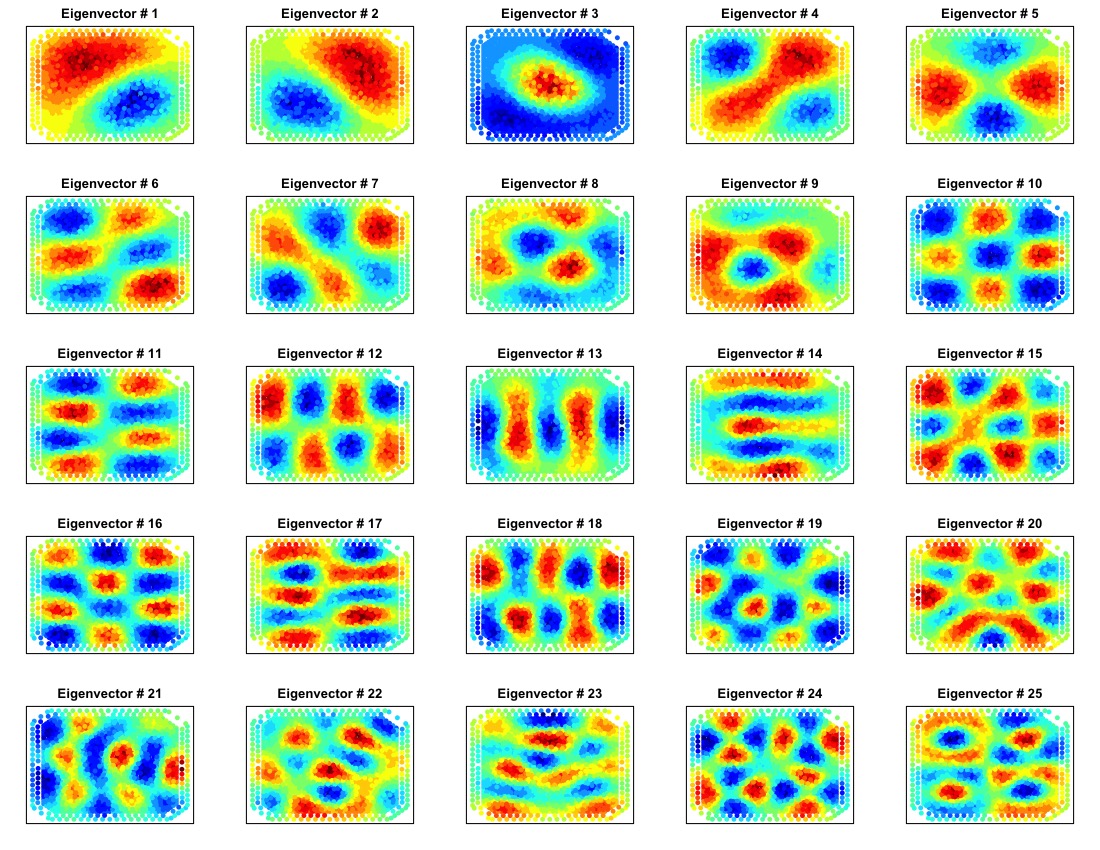}
    \caption{The leading 25 eigenvectors of the Moran's operator generated on the triangular mesh. The distinct spatial patterns construct the latent spatial random field for hierarchical spatial models.}
    \label{fig:eigencomponents}
\end{figure}

We construct the Moran's basis function matrix $\mathbf{M}\in \mathbb{R}^{m\times p}$, by selecting the first $p$ eigenvectors of the Moran's operator where $p\ll m$. The Moran's basis functions contains orthogonal spatial patterns generated on the triangular mesh vertices. Figure \ref{fig:eigencomponents} illustrates the first 25 eigenvectors of the Moran's operator. The first Moran's eigenvector contains the spatial pattern that corresponds to the largest positive value of the Moran's I test statistic (e.g. high level of spatial clustering) \citep{griffith2003spatial}. The subsequent eigenvectors are orthogonal and contain spatial patterns that correspond to smaller values of the Moran's I statistic (e.g. less spatial clustering). In Section \ref{SubSec:Tuning}, we provide an automated heuristic for selecting a suitable rank $p$. 

We generate a spatial random field on the mesh vertices by taking linear combinations of the Moran's basis functions (contained in matrix $\mathbf{M}$) and their corresponding weights $\pmb{\delta}\in \mathbb{R}^{p}$. Some of the eigenvectors contain spatial patterns with sharp peaks in localized regions. These ``rougher'' eigenvectors will be strongly down-weighted in our approximation of the latent spatial random field. Patterns that better-represent the true latent spatial random field will have higher magnitude weights while those that do not properly represent the underlying random field will be assigned near-zero weights. In Section \ref{SubSec:BayesianModel}, we provide a general framework for estimating $\pmb{\delta}$ in hierarchical spatial models.

\subsubsection*{Piece-wise Linear Basis Functions}
We introduce a set of piece-wise linear basis functions \citep{brenner2007mathematical} to interpolate points within the triangular mesh (i.e. the undirected graph $\bs{G}=(\bs{V},E)$). We construct a spatial random field $\mathbf{\tilde{W}}=(w(\bs{v}_{1}),...,w(\bs{v}_{m}))'$ on the mesh nodes $\bs{v}_{j}\in \bs{V}$. Then, we project, or interpolate, onto the observed locations $\bs{s}=\{\bs{s}_{1},...,\bs{s}_{n}\}$. The latent spatial random field $\mathbf{W}=(w(\bs{s}_{1}),...,w(\bs{s}_{n}))'$ at the observed locations can be represented as $\mathbf{W=A\tilde{W}}$, where $\mathbf{A}$ is an $n\times m$ projector matrix containing the piece-wise linear basis functions. 

The rows of $\mathbf{A}$ correspond to observation locations $\bs{s}_{i}\in \mathcal{D}$, and the columns correspond to mesh nodes $\bs{v}_{j}\in \bs{V}$. The $i$-th row of $\mathbf{A}$ contains the weights to linearly interpolate onto locations $\bs{s}_{i}$. In practice, we use an $n\times m$ projector matrix $\mathbf{A}$ for fitting the hierarchical spatial model. For model validation and prediction, we generate an $n_{CV} \times m$ projector matrix $\mathbf{A}_{CV}$ that interpolates onto the $n_{cv}$ validation locations.

The piece-wise linear basis functions can interpolate an observation location that is wholly contained in a triangle. Suppose point $\bs{s_0}$ is the location of interest, points $\bs{s}_1$, $\bs{s}_2$, and $\bs{s}_3$ are the triangle vertices, and $\pi_{1},\pi_{2},$ and $\pi_{3}$ are the weights, where $\sum_{i}^{3}\pi_{i}=1$. $\pi_{1}$ is the proportion of the area of the triangle opposite of vertex $\bs{s}_1$ to the area of the entire triangle. The same relationship holds for values $\pi_{2}$ and $\pi_{3}$ and vertices $\bs{s}_2$ and $\bs{s}_3$, respectively. We can approximate the value at point $\bs{s}_0$ as the weighted mean of the value observed at the three vertices $w(\bs{s}_{0})\approx \sum_{j=1}^{3}\pi_{j}w(\bs{s}_{j})$.

\subsection{Bayesian Hierarchical Spatial Model using PICAR}\label{SubSec:BayesianModel}
\begin{figure}[ht]
\centering
    \includegraphics[width=0.9\textwidth]{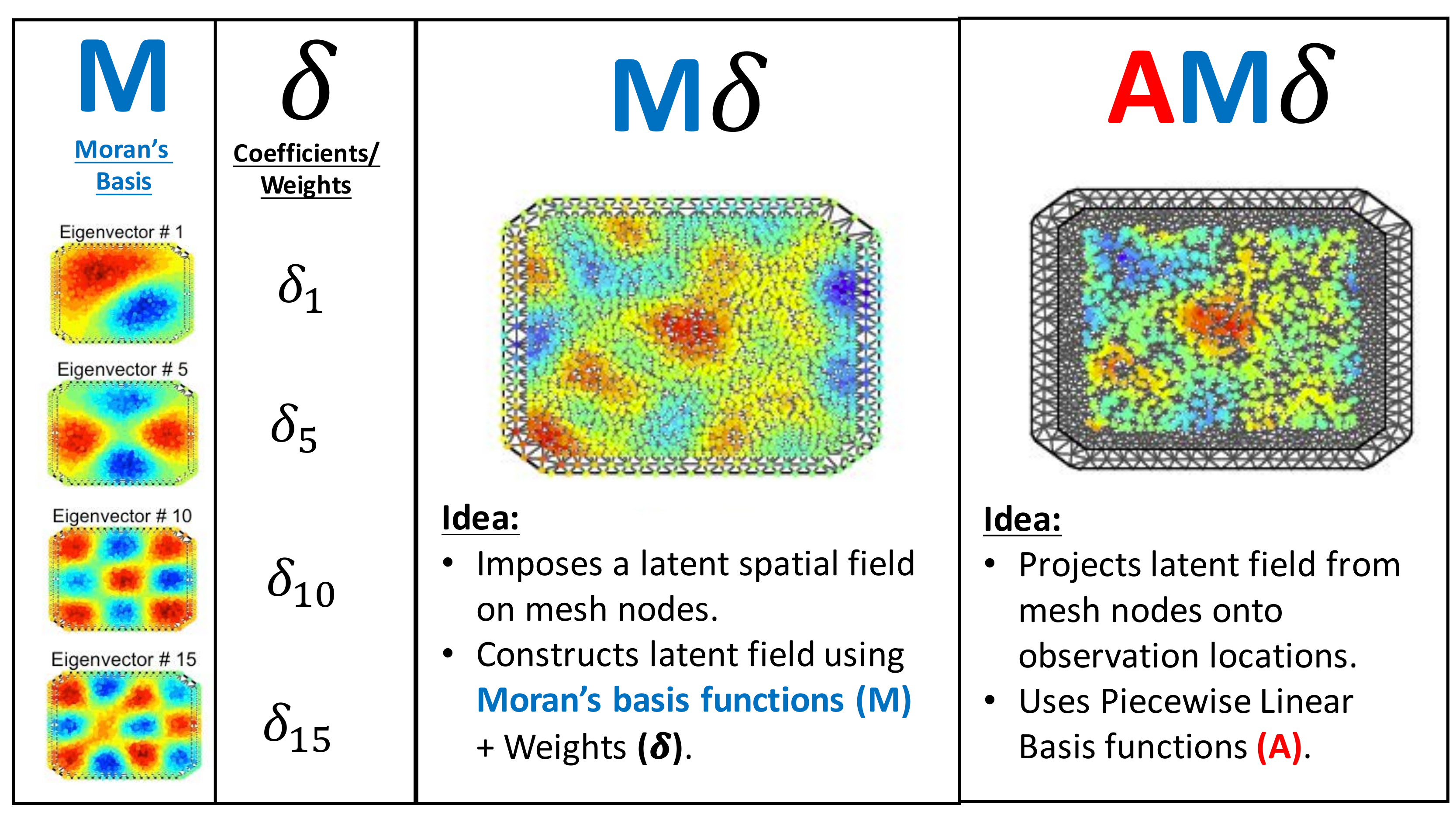}
    \caption{Diagram of the basis functions within PICAR. The Moran's basis functions (left) represent distinct spatial patterns, and the coefficients ($\pmb{\delta}$) denote the associated weights. The operation $\bf{M\pmb{\delta}}$ constructs a latent field on the mesh nodes. The operation $\mathbf{AM\pmb{\delta}}$ projects the mesh nodes onto the observation locations and generates a spatial random field.}
    \label{Fig:BasisRole}
\end{figure}

In the previous section, we introduced the three major components of PICAR: (1) the Moran's basis function matrix $\mathbf{M}\in \mathbb{R}^{m\times p}$; (2) the projector matrix $\mathbf{A}\in \mathbb{R}^{n\times m}$; and (3) the corresponding weights $\pmb{\delta} \in \mathbb{R}^{p}$. We can construct a spatial random field on the triangular mesh vertices as $\tilde{\mathbf{W}}=\mathbf{M}\pmb{\delta}$, where 
$\mathbf{\tilde{W}}=(w(\bs{v}_{1}),...,w(\bs{v}_{m}))'$ for $\bs{v}_{i}\in \bs{V}$. Next, we linearly interpolate, or approximate, the latent spatial random field at the observation locations as $\mathbf{W}\approx \mathbf{A}\mathbf{\tilde{W}}=\mathbf{AM}\pmb{\delta}$, where 
$\mathbf{W}=(w(\bs{s}_{1}),...,w(\bs{s}_{n}))'$ for $\bs{s}_{i}\in \mathcal{D}$. An overview of these operations is provided in Figure \ref{Fig:BasisRole}.

 PICAR can be embedded into the hierarchical spatial model framework as follows:
 \begin{align*}
     \mbox{\textbf{Data Model:}}&\qquad Z(\bs{s}_{i})|\mu(\bs{s}_{i}) \sim f(\mu(\bs{s}_{i})),\qquad \mu(\bs{s}_{i}) = \mathbb{E}[Z(\bs{s}_{i})|\pmb{\beta},\pmb{\delta}]\\
     &\qquad \eta(\bs{s}_{i}) = g(\mu(\bs{s}_{i})),\qquad \pmb{\eta}=(\eta(\bs{s}_1),...,\eta(\bs{s}_n))'\\
     &\qquad \pmb{\eta} = \mathbf{X}\pmb{\beta} + \mathbf{A}\mathbf{M}\pmb{\delta}\\
     \mbox{\textbf{Process Model:}}&\qquad \pmb{\delta}|\tau \sim \mathcal{N}(\bs{0}, \tau^{-1}(\mathbf{M'}\mathbf{Q}\mathbf{M})^{-1})\\
     \mbox{\textbf{Parameter  Model:}}&\qquad \pmb{\beta} \sim \mathcal{N}(\pmb{\mu}_{\pmb{\beta}},\pmb{\Sigma}_{\pmb{\beta}}),\quad \tau\sim \mathcal{G}(\alpha_{\tau},\beta_{\tau})
 \end{align*}
where $g(\cdot)$ is a link function, $\mathbf{A}$ is the projector matrix, $\mathbf{M}$ is the Moran's basis functions matrix, $\pmb{\delta}$ is the vector of basis coefficients, $\mathbf{Q}$ is the prior precision matrix for the mesh vertices, $\tau$ is the precision parameter, and $\alpha_{\tau},\beta_{\tau},\pmb{\mu}_{\pmb{\beta}},$ and $\pmb{\Sigma}_{\pmb{\beta}}$ are the model hyperparameters.

By default, we set $\mathbf{Q}$ to be the precision matrix of an intrinsic conditional auto-regressive model (ICAR) fit on the mesh vertices $\bs{V}$. Here, $\mathbf{Q}=(\mbox{diag}(\mathbf{N1})-\mathbf{N})$, where $\mathbf{N}$ is the adjacency or neighborhood matrix from Section \ref{SubSec:PICAR} and $\mathbf{1}$ is $m$-dimensional vector of 1s. Since $\mathbf{Q}$ is not positive definite, this framework cannot be used within the likelihood function; however, it can be set as the prior distribution for the spatial random effects as part of the Bayesian hierarchical spatial model \citep{besag1991bayesian}. We introduce alternative precision matrices in Section \ref{SubSec:Tuning} and provide a comparative analysis across matrices in Section \ref{Sec:SimulationStudy}. However, our simulation results indicate that prediction and inference is not sensitive to the choice of $\mathbf{Q}$.
 
The PICAR approach is flexible as it approximates the underlying spatial random field with a linear combination of the projected Moran's basis functions $\mathbf{AM}$ and basis coefficients $\pmb{\delta}$ such that $\mathbf{W}\approx\mathbf{AM}\pmb{\delta}$. The Moran's basis functions matrix $\mathbf{M}$ contains a rich collection of orthogonal spatial patterns located on the mesh vertices. The neighborhood, or weight, matrix $\mathbf{N}$ does not directly approximate the spatial field. Instead, we use the adjacency matrix $\mathbf{N}$ to generate the Moran's basis functions contained in $\mathbf{M}$. 

Suppose we have two latent spatial random fields $\{\mathbf{W}_{1},\mathbf{W}_{2}\}$ with differing spatial covariance structures. The PICAR approach would utilize a different set of Moran's basis functions $\{\mathbf{M}_{1},\mathbf{M}_{2}\}$ and basis coefficients $\{\pmb{\delta}_{1},\pmb{\delta}_{2}\}$ for each dataset such that  $\mathbf{W_{1}}\approx\mathbf{AM}_{1}\pmb{\delta}_{1}$ and $\mathbf{W_{2}}\approx\mathbf{AM}_{2}\pmb{\delta}_{2}$. To demonstrate the flexibility of the Moran's basis functions, we conducted numerical studies on simulated datasets (Section \ref{Sec:SimulationStudy}) that were generated using covariance functions from the Mat\'ern class. The PICAR approach performs comparably to the full hierarchical spatial model (gold standard) while incurring just a fraction of the computational cost.
    
The PICAR approach is subject to the following limitations: (1) may over-smooth the latent spatial random field; (2) spatial basis functions (Moran's eigenvectors) must be chosen before model-fitting; and (3) the user must set the appropriate mesh density. Since PICAR selects $p\ll n$, it is a low-rank approach that is subject to to oversmoothing \citep{Stein_2014}. However, simulation results suggest that oversmoothing may actually improve model performance for non-Gaussian observations, particularly binary and count data (Table \ref{Table:BinaryRank} and Supplement Tables 3,6, and 9). Models with lower rank $p$ (i.e. fewer basis functions) generally perform better than those with higher ranks. Though we provide an automated heuristic (Section \ref{SubSec:Tuning}) for basis selection and insights into mesh construction (Supplement Section 5), a comprehensive examination of these two areas would be beyond the scope of this study.

\subsection{Automating PICAR}\label{SubSec:Tuning}
The traditional hierarchical spatial model (Section \ref{SubSec:SGLMM}) assumes that the true latent spatial random field $\mathbf{W}$ is a Gaussian process such that $\mathbf{W} \sim \mathcal{N}(\bs{0},\bs{C}(\pmb{\Theta}))$ with covariance function parameters $\pmb{\Theta}$. On the other hand, PICAR considers the latent spatial random field following a basis representation such that  $\mathbf{W}\approx\mathbf{A}\mathbf{M}\pmb{\delta}$, where $\pmb{\delta}\sim \mathcal{N}(\bs{0},\tau^{-1}(\mathbf{M'Q}\mathbf{M})^{-1})$, $\mathbf{M}$ is the $m \times p$ Moran's basis function matrix, and $\mathbf{A}$ is the $n \times m$ projector matrix. An alternative formulation of the latent spatial random field is $\mathbf{W}\sim \mathcal{N}(0,\tau^{-1}\mathbf{A}\mathbf{M}(\mathbf{M'Q}\mathbf{M})^{-1}\mathbf{M'}\mathbf{A'})$.

Our objective is to accurately represent the true latent state using PICAR's basis representation. To that end, we can tune the rank of the Moran's operator $\mbox{rank}(\mathbf{M})$ and the prior precision matrix $\mathbf{Q}$ of the mesh vertices. The following automated heuristic selects an appropriate rank for the Moran's basis. First, we generate a set $\mathcal{P}$ consisting of $h$ equally spaced points within the interval $[2,P]$ where $P$ is the maximum rank and $h$ is the interval resolution ($h=P-1$ by default). Here, $P<m$ and both $P$ and $h$ are chosen by the user. For each $p\in\mathcal{P}$, we construct an $n\times(k+p)$ matrix of augmented covariates $\tilde{\bs{X}}=[\bs{X} \quad \mathbf{AM_{p}}]$ where $\bs{X}\in \mathbb{R}^{n\times k}$ is the original design matrix, $\mathbf{A} \in \mathbb{R}^{n\times m}$ is the projector matrix, and $\mathbf{M_{p}} \in \mathbb{R}^{m\times p}$ are the leading $p$ eigenvectors of the Moran's operator. Next, we use maximum likelihood approaches to fit the appropriate generalized linear model (GLM) for the response type (e.g. binary, count, or ordered categorical). Finally, we select the rank $p$ that yields the lowest out-of-sample cross-validated mean squared prediction error (CVMSPE). 

Next, we provide some choices for $\mathbf{Q}$, the prior precision matrix for the latent spatial field $\mathbf{\tilde{W}}$, located on the mesh vertices. By default (Section \ref{SubSec:PICAR}), we set $\mathbf{Q}$ to be the precision matrix of an intrinsic conditional auto-regressive model (ICAR). Similarly, we could set $\mathbf{Q}$ as the precision matrix of a conditional auto-regressive model (CAR). Here, $\mathbf{Q}=(\mbox{diag}(\mathbf{N1})-\rho \mathbf{N})$, where $\mathbf{N}$ is the adjacency matrix and $\rho\in(0,1)$ is a fixed correlation coefficient. It is possible to estimate $\rho$ as a model parameter, but doing so requires inverting a newly constructed matrix $\mathbf{M'QM}$ at each iteration of the MCMC algorithm; thereby increasing computational costs. Another alternative is setting $\mathbf{Q}=\bs{I}$, where $\mathbf{\tilde{W}}$ are uncorrelated. When $\mathbf{Q=I}$, the components of $\pmb{\delta}$ are also uncorrelated since the columns of $\bs{M}$ are orthogonal.

Rank selection, or the choice of $p$, varies based on the smoothness of the underlying spatial random field and the overall mesh density. Through a numerical study (Supplement Section 4) using count observations, we find that the automated heuristic selects a lower rank for data generated from smoother spatial random fields (larger $\nu$ in the case of the Mat\'ern class) than those from a `rougher' field. Results from our simulation study (Section \ref{Sec:SimulationStudy}) show that all three parameterizations of $\mathbf{Q}$ provide comparable inferential and prediction results. Moreover, all three parameterizations have a computational cost of $\mathcal{O}(np)$. However, we recommend using $\mathbf{Q}=\mathbf{I}$ as it leads to a faster-mixing MCMC algorithm with higher effective samples per second (ES/sec).

\subsection{Computational Gains}\label{SubSec:Comp}
\begin{figure}[ht]    
    \centering
    \includegraphics[width=0.8\textwidth]{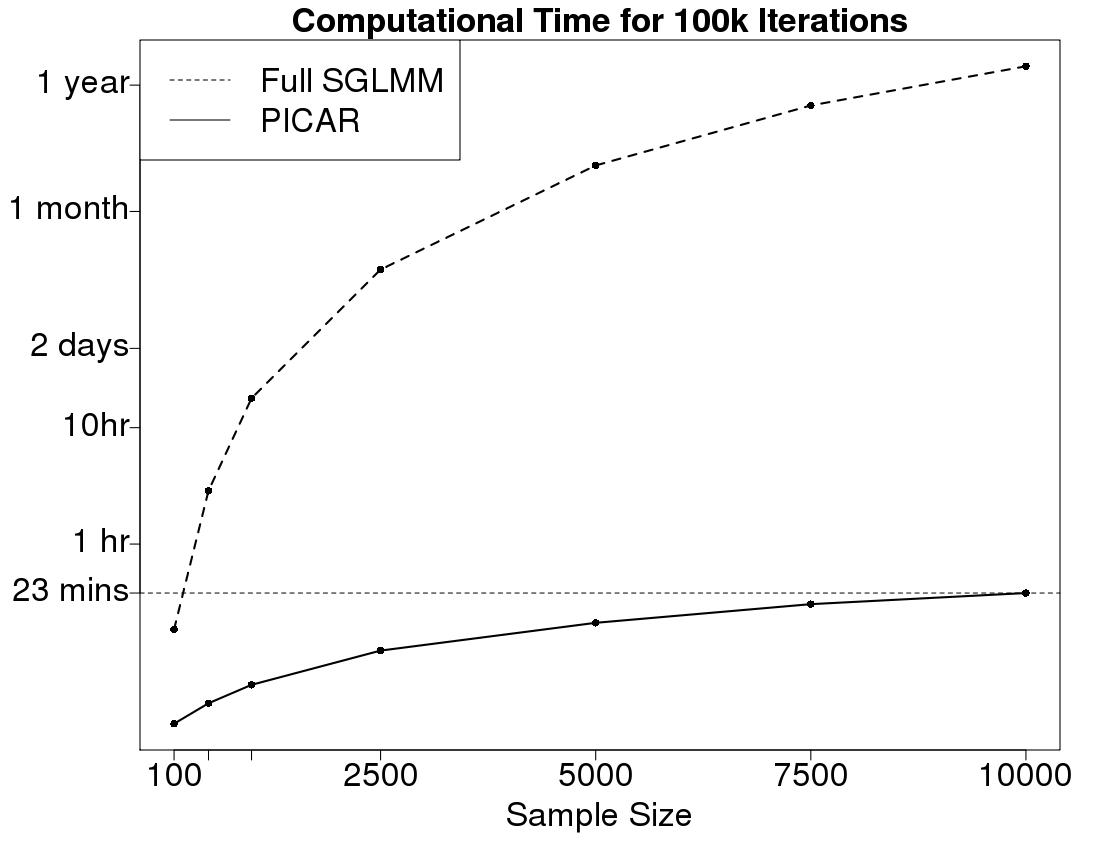}
    \caption{Computational time for $10^5$ iterations versus sample size ($n$) for the full spatial generalized linear mixed model and PICAR with chosen rank $p=50$.}
\label{fig:computationtime}
\end{figure}

PICAR incurs less computational costs per iteration as well as fewer iterations for the Markov chain to converge compared to the full hierarchical spatial model. The computational speedup results from bypassing expensive matrix operations (e.g. Cholesky decomposition) and by reducing the correlation and the dimensions of the spatial random effects. The computational cost is dominated by the matrix-vector multiplication $\mathbf{AM}\pmb{\delta}$, where $\textbf{AM}$ is the $n \times p$ basis function matrix constructed prior to model fitting and $\pmb{\delta}$ are reparameterized spatial random effects (basis coefficients). PICAR has a computational complexity of $\mathcal{O}(np)$ as opposed to $\mathcal{O}(n^{3})$ for the full hierarchical spatial model. Figure \ref{fig:computationtime} illustrates the computational speedup offered by PICAR. As we increase the dimensionality of the observations $n$, the full hierarchical model quickly becomes computationally prohibitive. On the other hand, we can fit the model using PICAR within the order of minutes. The computation times are based on a single 2.2 GHz Intel Xeon E5-2650v4 processor. All the code was run on the Pennsylvania State University Institute for CyberScience-Advanced CyberInfrastructure (ICS-ACI) high-performance computing infrastructure.

PICAR results in a faster mixing MCMC algorithm than the reparameterized MCMC algorithm (Rep-SGLMM) \citep{Christensen2006robust}, as shown by its larger effective samples per second (ES/sec), the rate at which independent samples are generated by the MCMC algorithm. In the binary data example, the PICAR ES/sec is roughly $345$ times larger than the Rep-SGLMM approach (details in Section \ref{SubSec:Binary}). 
We also show how to implement PICAR in the {\tt stan} \citep{carpenter2017stan} and \texttt{nimble} \citep{nimble2017} programming languages, which generate faster mixing MCMC algorithms.  

For PICAR, the two major computational bottlenecks are constructing the Moran's operator (Section \ref{SubSec:Moran}) and computing its eigencomponents. The Moran's operator requires the matrix operation $\mathbf{(I-11'/}m)\mathbf{N}(\mathbf{I-11'}/m)$ and $2m^{3}-m^{2}$ floating point operations (flops), which can be computationally prohibitive when $m$ is large (i.e. dense meshes). We can reduce computational costs by leveraging the embarrassingly parallel operations as well the sparsity of the adjacency matrix $\mathbf{N}$. Moreover, we can compute the leading $k$ eigencomponents of the Moran's operator using a partial eigendecomposition approach such as the Implicitly Restarted Arnoldi Method \citep{lehoucq1998arpack} from \textbf{RSpectra} package \citep{RSpectra2019}. Since the automated heuristic (Section \ref{SubSec:Tuning}) typically selects a rank($\bs{M}$) which is much lower than $m$, it is not necessary to perform a full eigendecomposition. 

\section{Simulation Study}\label{Sec:SimulationStudy}

We examine PICAR through simulation studies involving several hierarchical spatial models: (i) a binary spatial generalized linear mixed model shows that our approach produces similar results to fitting the full (``gold standard'') spatial model; (ii) a spatially varying coefficients model demonstrates how PICAR can be easily extended to user-specified models using the language {\tt stan} and {\tt nimble}; (iii) an ordered categorical data example shows how our approach is efficient for a problem for which other methods are not readily applicable. 

\subsection{Binary Data}\label{SubSec:Binary}

We generate $100$ datasets with randomly chosen locations on the unit square $\mathcal{D}=[0,1]^{2}$. Each dataset consists of $1,400$ locations allocating $1,000$ for model fitting and $400$ for validation. We chose $n=1,000$ to enable comparisons with fitting the full spatial generalized linear mixed model (SGLMM). The observations are generated from an SGLMM with a Bernoulli data model and a logit-link function $\mbox{logit}(p) = \log\{\frac{p}{1-p}\}$. We generate data from a hierarchical spatial model with regression coefficents $\pmb{\beta}=(1,1)'$ and Mat\'ern covariance function (Equation \ref{EQ:Matern}) parameters $\nu=2.5$, $\sigma^{2}=1$, and $\phi=0.2$. We compare PICAR's performance across varying ranks of the Moran's operator $p=\{10,50,75,100,200\}$. We also compare three different precision matrices $\bs{Q}$, Independent, ICAR, and CAR where the rank $p$ is chosen by our heuristic (Section \ref{SubSec:Tuning}). For the CAR precision matrix, we fixed the smoothing parameter $\rho=0.5$ prior to model fitting. Note that all three parameterizations of $\mathbf{Q}$ scales linearly. We compare PICAR's inference and prediction performance against a gold standard approach that uses the fast mixing MCMC algorithm in \citet{Christensen2006robust}. We set priors following  \citet{hughes2013dimension} where $\pmb{\beta}\sim \mathcal{N}(\mathbf{0},100\bs{I})$ and $\tau\sim \mathcal{G}(0.5,2000)$.

For PICAR, we use a triangular mesh with $m=1,649$ vertices.  In the binary case, we use Gibbs updates for $\tau$ and random-walk Metropolis-Hastings updates for $\pmb{\beta}$ with proposal $\pmb{\beta}^{(i+1)}\sim \mathcal{N}(\pmb{\beta}^{(i)},\hat{\bs{V}})$, where $\hat{\bs{V}}$ is the asymptotic covariance matrix from fitting the classical generalized linear model. Finally, we update the reparameterized random effects $\pmb{\delta}$ using an all-at-once normal random walk Metropolis-Hastings update \citep{Guan_Haran_2018}. We ran $300,000$ iterations of the MCMC algorithm for the PICAR approach. 

We select one sample (from the 100 generated samples) as the dataset for the comparative analysis. When comparing across ranks, we use the precision matrix from the ICAR model $\mathbf{Q}=(\mbox{diag}(\mathbf{N1})-\mathbf{N})$. We examine the out-of-sample cross-validated mean squared prediction error $\mbox{CVMSPE}=\frac{1}{n_{CV}}\sum_{i=1}^{n_{CV}}(Z(\bs{s}^{*}_{i})-\hat{Z}(\bs{s}^{*}_{i}))^{2},$ where $n_{CV}=400$, and $Z(\bs{s}^{*}_{i})$ and $\hat{Z}(\bs{s}^{*}_{i})$ denote the observed and predicted values at validation location $\bs{s}^{*}_{i}$, respectively. 

Table \ref{Table:BinaryRank} presents the parameter estimates, prediction results, and computational times for each rank $p$ of the Moran's operator. Results suggest that the rank is a key driver for predictive performance and parameter estimation. PICAR is not sensitive to the chosen precision matrix $\mathbf{Q}$, as the results are similar across precision matrices (Table 1 in Supplement). PICAR improves mixing in the MCMC algorithm as shown by the larger effective samples per second (ES/sec) compared to the gold standard approach. For model parameters $\beta_{1}$ and $\beta_{2}$, PICAR yields an ES/sec of $29.4$ and $40.2$ respectively and the gold standard returns an ES/sec $0.19$ and $0.29$ respectively. For the random effects $\mathbf{W}$, the average ES/sec is $5.8$ for the PICAR approach and $0.016$ for the gold standard, an improvement by a factor of roughly $345$.

\begin{table}[ht]
\centering
\begin{tabular}{lllrr}
  \hline
Rank & $\beta_{1}$ (95\% CI) & $\beta_{2}$ (95\% CI)& CVMPSE & Time (min) \\ 
  \hline
10 & 1.04 (0.77,1.31) & 0.91 (0.64,1.16) & 0.3 & 9.73 \\ 
  \textbf{22} & \textbf{1.09 (0.82,1.37)} & \textbf{0.93 (0.67,1.2)} & \textbf{0.27} & \textbf{10.73} \\ 
  50 & 1.12 (0.83,1.41) & 0.95 (0.67,1.23) & 0.28 & 11.14 \\ 
  75 & 1.14 (0.85,1.44) & 0.98 (0.69,1.26) & 0.28 & 11.62 \\ 
  100 & 1.2 (0.9,1.5) & 1 (0.71,1.29) & 0.29 & 12.28 \\ 
  200 & 1.34 (1.01,1.66) & 0.99 (0.69,1.31) & 0.32 & 15.13 \\ 
  Gold Standard & 1.03 (0.77,1.3) & 0.89 (0.63,1.16) & 0.29 & 3624.43 \\ 
   \hline
\end{tabular}

\caption{Simulated example with binary spatial observations. Parameter estimation, prediction, and model fitting time results across Moran's basis ranks. Bold font denotes the rank chosen by the automated heuristic.} 
\label{Table:BinaryRank}
\end{table}

Note that the PICAR approach is computationally efficient, and it also outperforms the gold standard approach in prediction. This is consistent with results from another basis representation approach, the latent conjugate model \citep{bradley2019bayesian}, which also outperforms the full SGLMM in computational cost and predictive ability. This may be attributed to the flexibility of PICAR's basis representation of the latent spatial field. 

We examine boxplots for the parameter estimates of $\beta_{1}$ and $\beta_{2}$ across the 100 samples (Supplement). The point estimates from the PICAR approach are distributed narrowly around the true values. The distribution of the point estimates remain similar across the choice of precision matrix $\mathbf{Q}$. The coverage proportions ($0.89$ for $\beta_{1}$, $0.91$ for $\beta_{2}$) are close to but lower than the nominal coverage value ($0.95$). 

\subsection{Count Data with Spatially Varying Coefficients}\label{SubSec:Poisson}
We incorporate the PICAR approach to the spatially varying coefficients model using both the {\tt stan} programming language \citep{carpenter2017stan} and \texttt{nimble} \citep{nimble2017}, two popular software environments for Bayesian inference. As in the binary example, we generate $n=1,400$ observations using the model parameters $\pmb{\beta}$ and $\phi$. We assign one set of spatially varying coefficients $\beta_{1}(\bs{s})$ corresponding to the first covariate $X_{1}(\bs{s})$ at location $\bs{s}$. $\mathbf{W}=(w(\bs{s}_{1}),...,w(\bs{s}_{n}))'$ and $\mathbf{B}=(\beta_{1}(\bs{s}_{1}),...,\beta_{1}(\bs{s}_{n}))'$ are the $n$-dimensional vector of the spatial random effects and spatially varying coefficients, respectively, for locations $\bs{s}_{i}\in \mathcal{D}$. It follows that:
$$\begin{bmatrix}
\bs{W}\\
\bs{B}
\end{bmatrix}\sim \mathcal{N}\Big(\mathbf{0},\bs{R}(\phi,\nu)\otimes \mathbf{T}\Big),\qquad \mathbf{T}=\begin{bmatrix}
1.0 & 0.3 \\
0.3 & 0.2 \\
\end{bmatrix}$$
where $\bs{R}(\phi,\nu)$ is the $n\times n$ Mat\'ern correlation matrix from the binary case with $\phi=0.2$ and $\nu=2.5$. In the PICAR framework, we represent the spatial processes  $\mathbf{B}$ and $\mathbf{W}$ by $\mathbf{B}\approx\mathbf{AM\pmb{\delta}_{\beta}}$ and $\mathbf{W}\approx\mathbf{AM\pmb{\delta}_{w}}$, where $\mathbf{A}$ is the $n\times m$ projector matrix, $\mathbf{M}$ is the $m\times p$ Moran's basis function matrix, and $\pmb{\delta}_{\beta}$ and $\pmb{\delta}_{w}$ are the $p$-dimensional vector of basis coefficients for the spatially varying coefficients and spatial random effects, respectively. We assume $\mathbf{B}$ and $\mathbf{W}$ are independent spatial processes with no cross-correlations. 

The PICAR specification of the spatially-varying coefficients model is as follows: 
\begin{align*}
\mbox{\textbf{Data Model:}}&\qquad Z(\bs{s}_{i})|\lambda(\bs{s}_{i}) \sim \mbox{Poisson}(\lambda(\bs{s}_{i})), \qquad \lambda(\bs{s}_{i})=\mathbb{E}[Z(\bs{s}_{i})|\pmb{\beta},\pmb{\delta}_{\beta},\pmb{\delta}_{w}]\\
    &\qquad   \eta(\bs{s}_{i}) = \log(\lambda(\bs{s}_{i})),\qquad \pmb{\eta}=(\eta(\bs{s}_1),...,\eta(\bs{s}_n))'\\
    &\qquad \pmb{\eta}=\mathbf{X\pmb{\beta}}+\bs{X}_{1}\odot\mathbf{AM\pmb{\delta}_{\beta}}+\mathbf{AM\pmb{\delta}_{w}}\\
    \mbox{\textbf{Process Model:}}& \qquad \pmb{\delta}_{\beta}\sim \mathcal{N}(\bs{0}, \tau_{\beta}^{-1}(\mathbf{M'}\mathbf{Q}\mathbf{M})^{-1}), \qquad \pmb{\delta}_{w}\sim \mathcal{N}(\bs{0}, \tau_{w}^{-1}(\mathbf{M'}\mathbf{Q}\mathbf{M})^{-1})\\
    \mbox{\textbf{Parameter  Model:}} &\qquad \tau_{\beta}\sim G(\alpha_{\tau1},\beta_{\tau1}),\quad \tau_{w}\sim G(\alpha_{\tau2},\beta_{\tau2}),\quad \pmb{\beta} \sim \mathcal{N}(\pmb{\mu}_{\pmb{\beta}},\pmb{\Sigma}_{\pmb{\beta}})
\end{align*}
where $\bs{X}_{1}$ is the first column of the $n \times 2$ design matrix $\mathbf{X}$, $\mathbf{Q}$ is the $m\times m$ precision matrix for the latent spatial field located on the mesh vertices, and $\tau_{\beta}$ and $\tau_{w}$ are the precision parameters. $\alpha_{\tau1},\beta_{\tau1},\alpha_{\tau2},\beta_{\tau2},\pmb{\mu}_{\pmb{\beta}},$ and $\pmb{\Sigma}_{\pmb{\beta}}$ are the hyperparameters. $\bs{X}_{1}\odot\mathbf{AM\pmb{\delta}_{\beta}}$ represents the Hadamard product, or component-wise multiplication, of the two $n$-dimensional vectors $\bs{X}_{1}$ and $\mathbf{AM\pmb{\delta}_{\beta}}$. In the process model, note that the $p-$dimensional vectors $\pmb{\delta}_{\beta}$ and $\pmb{\delta}_{w}$ replace the $n$-dimensional vectors \textbf{B} and \textbf{W} from the original model. 

We compare PICAR's performance across varying ranks $p=\{10,50, 63, 75,100,200\}$ of the Moran's operator. The lowest CVMSPE is achieved when we use the rank ($p=63$) selected via our automated heuristic. Model fitting times increase with respect to the chosen rank of the Moran's operator. Using {\tt stan}, we obtained a fast mixing MCMC algorithm, and an effective sample size of $\sim5,000$ for all parameters, random effects $\bs{W}$, and the spatially varying coefficients $\bs{B}$.   

For the spatially-varying coefficients model, we provide a comparative study with PICAR and three basis-representation approaches. Each approach represents the latent spatial random field using a class of spatial basis functions. We generate a simulated dataset of size $n=2800$ using the same model specifications presented above. The three competing basis functions are: (1) eigenvectors of a Mat\'ern covariance function with range parameter $\phi=0.2$, partial sill $\sigma^{2}=1$, and smoothness $\nu=2.5$; (2) bi-square basis function; and (3) thin-plate splines. Additional details regarding basis construction are available in the Supplement. The PICAR approach yields a lower cross-validated mean squared prediction error (CVMSPE) compared to the other basis representation methods. The bi-square basis functions and thin-plate splines exhibit faster mixing than PICAR or the Mat\'ern eigenvectors, as demonstrated by the higher effective samples per second (ESS). However, the former basis functions yield highly uncertain estimates of the model parameter $\beta_{1}$, in contrast to the PICAR or the Mat\'ern eigenvectors. The Mat\'ern eigenvectors requires fixing four tuning parameters, the range $\phi$, partial sill $\sigma^{2}$, smoothness $\nu$, and the rank of the basis functions matrix, prior to model fitting. Code for the comparative study is written in \texttt{nimble}. 

\subsection{Ordered Categorical Data}\label{SubSec:Ordered}
Let $Z(\bs{s})$ be the observation at $\bs{s}\in\mathcal{D}$ with $J$ ordered categories. We model ordered categorical data using the spatial cumulative-logit model through PICAR as follows:
\small
\begin{align*}
    \mbox{\textbf{Data Model:} }&\qquad  Z(\bs{s}_{i})|\pmb{\gamma}(\bs{s}_{i}) \sim \mbox{Cat}(J,\pmb{\pi}(\bs{s}_{i})),\qquad \pmb{\pi}(\bs{s}_{i})=(\pi_{1}(\bs{s}_{i}),...,\pi_{J}(\bs{s}_{i}))'=f(\pmb{\gamma}(\bs{s}_{i}))\\
    &\qquad \pmb{\gamma}(\bs{s}_{i})=(\gamma_{1}(\bs{s}_{i}),...,\gamma_{J}(\bs{s}_{i}))'\\
        & \qquad \gamma_{j}(\bs{s}_{i})|\pmb{\beta},\theta_{j},w(\bs{s}_{i})=\frac{\text{exp}\{\theta_{j}-(\mathbf{X}^{\prime}(\bs{s}_{i})\pmb{\beta}+w(\bs{s}_{i}))\}}{1+\text{exp}\{\theta_{j}-(\mathbf{X}^{\prime}(\bs{s}_{i})\pmb{\beta}+w(\bs{s}_{i}))\}}, \quad j=1,..,J-1 \\
    & \qquad \theta_{j}|\pmb{\alpha}=\sum_{i=1}^{j-1}\exp\{\alpha_{i}\},\qquad \pmb{\alpha}=(\alpha_{1},\alpha_{2},...,\alpha_{J-1})'\\
    &\qquad  \bs{W}=\mathbf{AM}\pmb{\delta}, \qquad \bs{W}=(w(\bs{s}_{1}),...,w(\bs{s}_{n}))'\\
    \mbox{\textbf{Process  Model:}} & \qquad \pmb{\delta}|\tau \sim \mathcal{N}(\bs{0}, \tau^{-1}(\mathbf{M'}\mathbf{Q}\mathbf{M})^{-1}),\\
    \mbox{\textbf{Parameter Model:}}&\qquad \pmb{\alpha}\sim p(\pmb{\alpha}),\quad \pmb{\beta} \sim \mathcal{N}(\pmb{\mu}_{\beta},\pmb{\Sigma}_{\pmb{\beta}}),\quad \tau\sim \mathcal{G}(\alpha_{\tau},\beta_{\tau})
\end{align*}
\normalsize
where $\pi_{j}(\bs{s}_{i})$, $\gamma_{j}(\bs{s}_{i})$ and $\theta_j(\bs{s}_{i})$ are the associated probability, cumulative probability and intercept (``cutoff") for the $j$-th category at the $i$-th location, respectively. The function $f(\cdot)$ transforms the cumulative probabilities to the associated probabilities (provided in the Supplement). $\mathbf{A}$ is the $n\times m$ projector matrix, $\mathbf{M}$ is the $m\times p$ Moran's basis functions matrix, $\pmb{\delta}$ is the $p$-dimensional vector of basis coefficients, and $\pmb{\mu}_{\pmb{\beta}}$, $\pmb{\Sigma}_{\pmb{\beta}}$, $\alpha_{\tau}$, and $\beta_{\tau}$ are hyperparameters.

We fix the first cutoff $\theta_{1}=0$ to avoid identifiability issues \citep{johnson2006ordinal}. Note that the $\theta_{j}$'s are constrained by the ordering  $\theta_{j}>\theta_{k}$ for $j>k$. Through a transformation \citep{higgs2010clipped,chib1997bayesian}, we can generate unconstrained cutoff parameters $\pmb{\alpha}=(\alpha_{1},\alpha_{2},...,\alpha_{J-1})$, where $\alpha_{1}=-\infty$, $\alpha_{2}=\log(\theta_{2})$, and $\alpha_{3}=\log(\theta_{3}-\theta_{2})$. Additional details on the spatial cumulative-logit model is provided in the Supplement. 

We generate 100 samples with the observations categorized into $J=4$ groups. The true cut-off parameters are $\theta_{1}=0$, $\theta_{2}=1$ and $\theta_{3}=2$. The other model parameters are similar to that in the binary case. To assess predictive performance, we examine the out-of-sample misprediction rate (MPR), or the proportion of incorrect predictions, and the loss function is $\mbox{MPR}=\frac{1}{n_{CV}}\sum_{i=1}^{n_{CV}}I(Z^{*}(\bs{s}_{i})\neq\hat{Z}^{*}(\bs{s}_{i})),$ for the observed $Z(\bs{s}^{*}_{i})$ are predicted values $\hat{Z}(\bs{s}^{*}_{i})$ at validation location $\bs{s}^{*}_{i}$. 

The automated heuristic chose a rank of $p=23$, which yields comparable parameter estimation and predictive ability to the gold standard. Predictive ability does not vary considerably across rank, but  rank impacts parameter estimation. PICAR is not sensitive to the chosen precision matrix $Q$, as the inferential and predictive performances do not vary across precision matrices. Similar to the binary case, we observe faster mixing with the PICAR approach: for $\beta_{1}$ and $\beta_{2}$, PICAR has an ES/sec of $30.7$ and $30.4$, while the gold standard has an ES/sec $0.034$ and $0.034$ respectively. For the random effects, PICAR has an average ES/sec of $4.4$ versus  $0.003$ for the gold standard, a speedup by a factor of around $1,487$. For the simulation study, we examine boxplots for the parameter estimates for $\beta_{1}$, $\beta_{2}$, $\alpha_{1}$, and $\alpha_{2}$ across all 100 samples (Supplement). Similar to the binary case, our coverage proportions are very close ($0.91$ for $\beta_{1}$, $0.92$ for $\beta_{2}$, $0.93$ for $\alpha_{1}$,  $0.88$ for $\alpha_{2}$), but slightly lower than the nominal coverage ($0.95$).

\section{Real Data Examples}\label{Sec:Applications}

We demonstrate how PICAR scales well to high-dimensional spatial datasets. 

\subsection{Binary Data: MODIS Cloud Mask Data} \label{SubSec:MODIS}
The National Aeronautics and Space Administration (NASA) launched the Terra Satellite in December 1999 as the flagship mission of the Earth Observing System. As in past studies \citep{sengupta2013hierarchical, bradley2019bayesian}, we examine cloud mask captured by the Moderate Resolution Imaging Spectroradiometer (MODIS) instrument on board the Terra satellite. The response is a binary incidence of cloud mask at a 1km $\times$ 1km spatial resolution at $n=2,748,620$ locations. In this study, we selected $2,473,758$ observations to fit our model and reserved $10\%$ of the total observations for validation. We include the vector $\mathbf{1}$ and a vector of latitudes as the covariates, similar to previous studies \citep{sengupta2013hierarchical, bradley2019bayesian}. We construct a triangular mesh with $m=7,910$ vertices, and our automated heuristic (Section \ref{SubSec:Tuning}) selected a rank of $p=150$ for the Moran's basis functions. The PICAR approach required $66$ hours to run $10^{5}$ iterations of the MCMC algorithm. Due to the computational costs, comparison with the full SGLMM is computationally infeasible. For classification, we set the threshold of the posterior predictive probabilities to be $0.38$, which is approximately the midpoint of the entire range of probabilities.  The false positive rate is $0.19$ and the false negative rate is $0.25$. These are comparable to the results in \citet{bradley2019bayesian}, which reports a false positive rate of $0.22$ and a false negative rate of $0.28$. We implement the PICAR approach in the probabilistic programming language \texttt{nimble} \citep{nimble2017}.

\subsection{Ordered Categorical Data: Maryland Stream Waders}\label{SubSec:MarylandStreamWaders}
Beginning in 2000, the Maryland Stream Waders (MSW) program is a statewide volunteer stream monitoring program managed by the Maryland Department of Natural Resources (DNR) and the Monitoring and Non-Tidal Assessment Division (MANTA). For each sample, the DNR laboratory calculated a Benthic Index of Biotic Integrity (BIBI) on a scale from 1 to 5. Each site was rated as Good (4-5), Fair (3-3.9), or Poor (1-2.9) \citep{stribling1998development}. A total of $6,951$ samples were collected within a 17-year time period (2000-2017) at irregular sampling locations \citep{MSW2019}. We fit the model using $5,561$ randomly selected observations and validate the model with the remaining $1,390$ samples. The observations are modeled using a spatial cumulative-logit model (Section \ref{SubSec:Ordered}) with just an intercept term. We generated a mesh with $m=8,810$ nodes and the automated heuristic chose a rank of $p=653$. The time to fit the spatial cumulative-logit model via PICAR is around 35 minutes. We estimate fitting the full model would require on the order of weeks to provide similarly accurate inference. 

\section{Discussion}\label{Sec:Discussion} 
In this study, we propose a fast extendable projection-based approach (PICAR) for modeling a wide range of hierarchical spatial models. In cases where it is possible to fit the full hierarchical spatial model, we show that our approach yields comparable results in both inference and prediction. We also provide a variety of other examples that illustrate the flexibility of the PICAR approach as well as the ease with  which non-experts can specify and efficiently fit their own hierarchical spatial models. We show that our approach is computationally efficient, scales up to higher dimensions, automated, and extendable to a variety of hierarchical spatial models. We provide an example of a hierarchical spatial model (ordinal spatial data) that cannot be fit using existing publicly available code but can be easily fit using PICAR. Moreover, we show that our approach is amenable to implementation in programming languages for Bayesian inference ({\tt stan} and {\tt nimble}). As shown in our real-data applications, our approach scales well to higher dimensions (up to $n=2.7$ Million). Where other approaches may be computationally infeasible, we can fit a high-dimensional hierarchical spatial model within hours. 

For larger spatial datasets ($n\geq1000$) with non-Gaussian observations, we recommend the PICAR model over the SGLMM (gold standard). PICAR provides a notable reduction in computational costs, yet provides comparable inference and prediction. For smaller datasets ($n<1000$), we would recommend fitting an SGLMM using the reparameterization approach \citep{Christensen2006robust} as model-fitting can be completed in a reasonable amount of time (order of hours). For Gaussian observations, state-of-the-art methods such as nearest-neighbor Gaussian processes, multiresolution approaches, or covariance tapering may be more or less efficient than PICAR, as mentioned in Section \ref{Sec:Challenges}.

The computational complexity for the PICAR approach is driven by matrix-vector multiplications, which can be readily parallelized. With efficient parallelization methods, we expect our approach to scale up to tens of millions of data points. Even though an eigendecomposition is only carried out once in our approach, methods such as Nystr\"{o}m method \citep{Williams01usingthe} or random projections \citep{Banerjee2013, Guan_Haran_2018} can further reduce costs via an approximate eigendecomposition of the Moran's operator. There may be other methods to improve our automated heuristic for rank selection such as a screening process for the relevant basis functions via a variable selection approach like LASSO \citep{Tibshirani94regressionshrinkage}. Assigning shrinkage priors on the basis coefficients \citep{bhattacharya2011sparse, legramanti2020bayesian} can better quantify the uncertainty attributed to the choice of basis functions. Extending the PICAR approach to spatio-temporal or multivariate spatial processes as well as computer model calibration with non-Gaussian model outputs may provide fruitful avenues for future research.

\section{Acknowledgments}\label{Sec:Acknowledgements}
We are grateful to Ephraim Hanks, Erin Schliep, Yawen Guan, Jaewoo Park, and Klaus Keller for helpful discussions. We would like to thank Perry de Valpine and Chris Paciorek for their assistance with \texttt{nimble}, particularly with the massive cloud mask dataset . The research was partially supported by the National Institute of General Medical Sciences of the National Institutes of Health under Award Number R01GM123007. The content is solely the responsibility of the authors and does not necessarily represent the official views of the National Institutes of Health. This work was partially supported by the U.S. Department of Energy, Office of Science, Biological and Environmental Research Program, Earth and Environmental Systems Modeling, MultiSector Dynamics, Contract No. DE-SC0016162 and by the National Science Foundation through the Network for Sustainable Climate Risk Management (SCRiM) under NSF cooperative agreement GEO-1240507. This study was also co-supported by the Penn State Center for Climate Risk Management. Any opinions, findings, and conclusions or recommendations expressed in this material are those of the authors and do not necessarily reflect the views of the Department of Energy, the National Science Foundation, or other funding entities. Any errors and opinions are those of the authors. We are not aware of any real or perceived conflicts of interest for any authors.

\bigskip
\begin{center}
{\large\bf SUPPLEMENTARY MATERIAL}
\end{center}

\begin{description}

\item[Supplemental Descriptions, Tables, and Figures:] Descriptions of various hierarchical spatial models, a parallelized method to generate the Moran's basis functions, three other simulation studies, and a real-world application. Additional figures and tables for the simulation studies, the dwarf mistletoe study, and the Maryland Stream Waders application (.pdf file)

\item [Code for the PICAR approach in \texttt{nimble} and \texttt{stan}:] Code to implement PICAR for various hierarchical spatial models can be accessed at the following github repository: \url{https://github.com/benee55/PICAR_code}. (github respository link)

\item[Dwarf mistletoe data set:] Data set used in the dwarf mistletoe application. (.csv file)

\item[Maryland Stream Waders data set:] Data set used in the Maryland Stream Waders application (Section \ref{SubSec:MarylandStreamWaders}). (.csv file)

\end{description}

\bibliographystyle{apalike}
\bibliography{references}
\end{document}


\doublespacing
\maketitle

\section{Examples of Hierarchical Spatial Models}\label{Sec:Other}
Here we provide examples of hierarchical spatial models. The first is the class of spatial generalized linear mixed models for non-Gaussian data, the second is a spatially-varying coefficients model, and the third is a cumulative-logit model for ordered categorical data.

\subsection*{Spatial Generalized Linear Mixed Models}
Non-Gaussian spatial observations are typically modeled using spatial generalized linear mixed models (SGLMMs) (Diggle et al., 1998, Haran, 2011). Let $\{Z(\bs{s}):\bs{s}\in \mathcal{D}\}$ be a non-Gaussian spatial random field. Assuming $Z(\bs{s})$ are conditionally independent given the latent random spatial random effect $w(\bs{s})$, the conditional mean $E[Z(\bs{s})|\pmb{\beta},w(\bs{s}),\epsilon(\bs{s})]$ can be modeled through the linear predictor $\eta(\bs{s})$:

$$\eta(\bs{s})=g\{E[Z(\bs{s})|\pmb{\beta},w(\bs{s}),\epsilon(s)]\}=\bs{X}^{\prime}(\bs{s})\pmb{\beta} +w(\bs{s})+\epsilon(\bs{s}),$$
where $g(\cdot)$ is a known link function. Binary and count observations are two common types of non-Gaussian spatial data, and these can be modeled using the binary SGLMM with logit link and the Poisson SGLMM with log link, respectively. 

The general Bayesian hierarchical framework for the SGLMM is:
\begin{align*}
    \mbox{Data Model: }&\qquad Z(\bs{s}_{i})|\mu(\bs{s}_{i}) \sim f(\mu(\bs{s}_{i})), \qquad \mu(\bs{s}_{i})=\mathbb{E}[Z(\bs{s}_{i})|\pmb{\beta},w(\bs{s}_{i}),\epsilon(\bs{s}_{i})]\\
    & \qquad \eta(\bs{s}_{i}) = g(\mu(\bs{s}_{i}))= \bs{X}^{\prime}(\bs{s}_{i})\pmb{\beta} + w(\bs{s}_{i}) + \epsilon(\bs{s}_{i})\\
    \mbox{Process  Model: } & \qquad \mathbf{W}|\pmb{\Theta} \sim N(\bs{0},\bs{C}(\pmb{\Theta})), \qquad \mathbf{W}=(w(\bs{s}_{1}),..,w(\bs{s}_{n}))'\\
    & \qquad \epsilon(\bs{s}_{i})|\tau^{2} \sim \mathcal{N}(0, \tau^{2})\\
    \mbox{Parameter  Model: } & \qquad  \pmb{\beta}\sim p(\pmb{\beta}),\quad \pmb{\Theta}\sim p(\pmb{\Theta}),\quad \tau^{2}\sim p(\tau^{2})
\end{align*}
where $f(\cdot)$ is a probability distribution (e.g. Bernoulli or Poisson), $\pmb{\beta}$ and $\bs{X}(\bs{s})$ are the $k$-dimensional vectors of the fixed effects and covariates for location $\bs{s}_{i}$, respectively. $\mathbf{W}$ is the $n$-dimensional vector of the spatial random effects, and  $\tau^{2}$ is the nugget variance. $\bs{C}(\pmb{\Theta})$ is the covariance matrix for the spatial random effects with covariance parameter $\pmb{\Theta}$.

\subsection*{Spatially Varying Coefficient Models}
Spatially varying coefficient models \citep{gelfand2003spatial} consider cases where the fixed effects $\pmb{\beta}$ vary across space. For the case of Gaussian responses $Z(\bs{s})$ with a single predictor $X(\bs{s})$, the data model is $Z(\bs{s})=\beta_{0}+\beta_{1}X(\bs{s})+\beta_{1}(\bs{s})X(\bs{s})+w(\bs{s})+\epsilon(\bs{s})$, where  $\beta_{0}$ is the intercept, $\beta_{1}$ is the fixed effect, $\beta_{1}(\bs{s})$ is the spatially varying coefficient term, and  $w(\bs{s})$ and $\epsilon(\bs{s})$ are the spatial random effects and micro-scale measurement errors, respectively. $\mathbf{B}=(\beta_{1}(\bs{s}_1),...,\beta_{1}(\bs{s}_n))'$ is the $n$-dimensional vector of spatially varying coefficients, and $\mathbf{B}\sim \mathcal{N}(\bs{0},\bs{C}_{\pmb{\beta}}(\pmb{\Theta}_{\pmb{\beta}}))$ where $\bs{C}_{\pmb{\beta}}(\pmb{\Theta})$ is the covariance matrix and $\pmb{\Theta}_{\pmb{\beta}}$ is the covariance parameter for the spatial random process $\mathbf{B}$. We model $w(\bs{s})$ and $\epsilon(\bs{s})$ as in the SGLMM case.

For cases with $k$ predictors, we have the following hierarchical spatial model:
\small
\begin{align*}
\mbox{\textbf{Data Model:}}&\qquad Z(\bs{s}_i)|\lambda(\bs{s}_i) \sim \mbox{Poisson}(\lambda(\bs{s}_i)), \qquad \lambda(\bs{s}_i)=\mathbb{E}[Z(\bs{s}_i)|\pmb{\beta}, \pmb{\beta}(\bs{s}_i),w(\bs{s}_i), \epsilon(\bs{s}_i)]\\
    &\qquad   \eta(\bs{s}_i) = \log(\lambda(\bs{s}_i))= \bs{X}^{\prime}(\bs{s}_i)\pmb{\beta}+\bs{X}^{\prime}(\bs{s}_i)\pmb{\beta}(\bs{s}_i)+w(\bs{s}_i)+\epsilon(\bs{s}_i)\\
    \mbox{\textbf{Process  Model:}} & \qquad \bs{B}|\pmb{\Theta}_{\pmb{\beta}},\mathbf{T} \sim \mathcal{N}(\bs{0},\bs{C}_{\pmb{\beta}}(\pmb{\Theta}_{\pmb{\beta}})\otimes \mathbf{T}), \qquad \mathbf{B}=(\pmb{\beta}^{\prime}(\bs{s}_{1}),...,\pmb{\beta}^{\prime}(\bs{s}_{n}))'\\
    & \qquad \bs{W}|\pmb{\Theta}_{W} \sim \mathcal{N}(\bs{0}, \bs{C}_{W}(\pmb{\Theta}_{W})), \qquad \bs{W}=(w(\bs{s}_{1}),...,w(\bs{s}_{n}))'\\
& \qquad \epsilon(\bs{s})|\tau^{2} \sim \mathcal{N}(0, \tau^{2})\\
\mbox{\textbf{Parameter  Model:}} & \qquad \pmb{\beta}\sim\pi(\pmb{\beta}),\quad \tau^{2}\sim\pi(\tau^{2}),\quad \pmb{\Theta}_{\pmb{\beta}}\sim\pi(\pmb{\Theta}_{\pmb{\beta}}),\quad \pmb{\Theta}_{W}\sim\pi(\pmb{\Theta}_{W}),\quad \mathbf{T}\sim\pi(\mathbf{T})
\end{align*}
\normalsize
where $\pmb{\beta}$ and $\bs{X}(\bs{s}_{i})$ are respectively, the $(k+1)$-dimensional vectors of the fixed effects and covariates for location $\bs{s}_{i}$. $\mathbf{B}$ is the $nk$-dimensional vector of all spatially varying coefficients. Note that we do not include the intercept in $\mathbf{B}$ as the spatially varying intercept term would confound with the spatial random effects $w(\bs{s})$. $\mathbf{W}$ is the $n$-dimensional vector of the spatial random effects, and  $\tau^{2}$ is the nugget variance. $\bs{C}_{\pmb{\beta}}(\pmb{\Theta}_{\pmb{\beta}})$ is the covariance matrix for the spatially varying coefficients with covariance parameter $\pmb{\Theta}_{\pmb{\beta}}$ and $\bs{C}_{W}(\pmb{\Theta}_{W})$ is the covariance matrix for the spatial random effects with covariance parameter $\pmb{\Theta}_{W}$.
$\mathbf{T}$ is a $k \times k$ positive definite matrix containing the marginal variances of $\pmb{\beta}_{j}(\bs{s})$'s on the diagonal elements and the cross-covariances on the off-diagonal elements for $j=1,...,k$.

\subsection*{Cumulative-Logit Models for Ordinal Spatial Data}
Ordered categorical (ordinal) data are categorical responses with a natural ordering, and commonly used in survey questionnaires, patient responses in clinical trials, and  quality assurance ratings for industrial processes. \citep{higgs2010clipped,schliep2013multilevel} develop a hierarchical spatial model for ordinal data. In this study, we examine the proportional-odds cumulative logit model \citep{agresti2010analysis} for ordered categorical data. Let $Z(\bs{s})$ be the observations at location $\bs{s}\in\mathcal{D}$ with $J$ ordered categories. Note that each ordered category corresponds to a probability $\pmb{\pi}(\bs{s})=(\pi_{1}(\bs{s}),\pi_{2}(\bs{s}),...,\pi_{J}(\bs{s}))$ , where $\pi_{i}(\bs{s})=\mbox{Pr}(Z(\bs{s})=i)$ for $i=1,...,J$. Here, we consider $J-1$ cumulative probabilities denoted as $\gamma_{j}(\bs{s})=P(Z(\bs{s})\leq j)=\pi_{1}(\bs{s})+...+\pi_{j}(\bs{s})$. The cumulative logit is defined as: 

$$\text{log}\left(\dfrac{P(Z(\bs{s}) \leq j)}{1-P(Z(\bs{s}) \leq j)}\right)=\text{log}\left(\dfrac{\gamma_{j}(\bs{s})}{1-\gamma_{j}(\bs{s})}\right)=\theta_j-\bs{X}^{\prime}(\bs{s})\pmb{\beta}-w(\bs{s})-\epsilon(\bs{s}),$$
where $\theta_j$ is the intercept or ``cutoff" for the $j$-th category, $\bs{X}(\bs{s})$, $\pmb{\beta}$, $w(\bs{s})$ and $\epsilon(\bs{s})$ are the spatial random effects and micro-scale measurement errors. The model for the cumulative probabilities $\gamma_{j}$ is: 
$$\gamma_{j}(\bs{s})=P(Z(\bs{s}) \leq j)=\frac{\text{exp}\{\theta_{j}-(\bs{X}^{\prime}(\bs{s})\pmb{\beta}+w(\bs{s})+\epsilon(\bs{s}))\}}{1+\text{exp}\{\theta_{j}-(\bs{X}^{\prime}(\bs{s})\pmb{\beta}+w(\bs{s})+\epsilon(\bs{s}))\}}.$$

Consequently, the probabilities for the individual $J$ categories are: 

\[P(Z(\bs{s})=j) = \left\{
  \begin{array}{ll}
    \gamma_{1}(\bs{s}), & \qquad j=1\\
    \gamma_{j}(\bs{s})-\gamma_{j-1}(\bs{s}), & \qquad 2\leq j \leq J-1\\
    1-\gamma_{J-1}(\bs{s}), & \qquad j=J\\
  \end{array}
\right.
\]

To avoid identifiability issues, we typically fix the first cutoff to be $\theta_{1}=0$ \citep{johnson2006ordinal}. Note that the $\theta_{j}$'s are constrained by the ordering  $\theta_{j}>\theta_{k}$ for $j>k$. Through a transformation \citep{higgs2010clipped,chib1997bayesian}, we can generate unconstrained cutoff parameters $\pmb{\alpha}=(\alpha_{1},\alpha_{2},...,\alpha_{J-1})$, where $\alpha_{1}=-\infty$, $\alpha_{2}=\log(\theta_{2})$, and $\alpha_{j}=\log(\theta_{j}-\theta_{j-1})$ for $j=3,...,J-1$. The inverse transformation is given by $\theta_{j}=\sum_{i=2}^{j}\mbox{exp}\{\alpha_{i}\}$. The hierarchical spatial model framework is as follows:

\small
\begin{align*}
    \mbox{\textbf{Data Model:} }&\qquad  Z(\bs{s}_{i})|\pmb{\gamma}(\bs{s}_{i}) \sim \mbox{Cat}(J,\pmb{\pi}(\bs{s}_{i})),\qquad \pmb{\pi}(\bs{s}_{i})=(\pi_{1}(\bs{s}_{i}),...,\pi_{J}(\bs{s}_{i}))'=f(\pmb{\gamma}(\bs{s}_{i}))\\
    &\qquad \pmb{\gamma}(\bs{s}_{i})=(\gamma_{1}(\bs{s}_{i}),...,\gamma_{J}(\bs{s}_{i}))'\\
        & \qquad \gamma_{j}(\bs{s}_{i})|\pmb{\beta},\theta_{j},w(\bs{s}_{i}),\epsilon(\bs{s}_{i})=\frac{\text{exp}\{\theta_{j}-(\mathbf{X}^{\prime}(\bs{s}_{i})\pmb{\beta}+w(\bs{s}_{i})+\epsilon(\bs{s}_{i}))\}}{1+\text{exp}\{\theta_{j}-(\mathbf{X}^{\prime}(\bs{s}_{i})\pmb{\beta}+w(\bs{s}_{i})+\epsilon(\bs{s}_{i}))\}}, \quad j=1,..,J-1 \\
    & \qquad \theta_{j}|\pmb{\alpha}=\sum_{i=1}^{j-1}\exp\{\alpha_{i}\},\qquad \pmb{\alpha}=(\alpha_{1},\alpha_{2},...,\alpha_{J-1})'\\
    \mbox{\textbf{Process  Model:}} & \qquad \bs{W}|\pmb{\Theta} \sim \mathcal{N}(\bs{0},\bs{C}(\pmb{\Theta})),\qquad \bs{W}=(w(\bs{s}_{1}),...,w(\bs{s}_{n}))'\\
    & \qquad \epsilon(\bs{s}_{i})|\tau^{2} \sim \mathcal{N}(0, \tau^{2}),\\
    \mbox{\textbf{Parameter Model:}}&\qquad \pmb{\alpha}\sim p(\pmb{\alpha}),\quad \pmb{\beta}\sim p(\pmb{\beta}),\quad \pmb{\Theta}\sim p(\pmb{\Theta}),\quad \tau^{2}\sim p(\tau^{2})
\end{align*}
\normalsize

\begin{figure}\label{Fig:Mesh}
    \centering
    \includegraphics[width=0.85\textwidth]{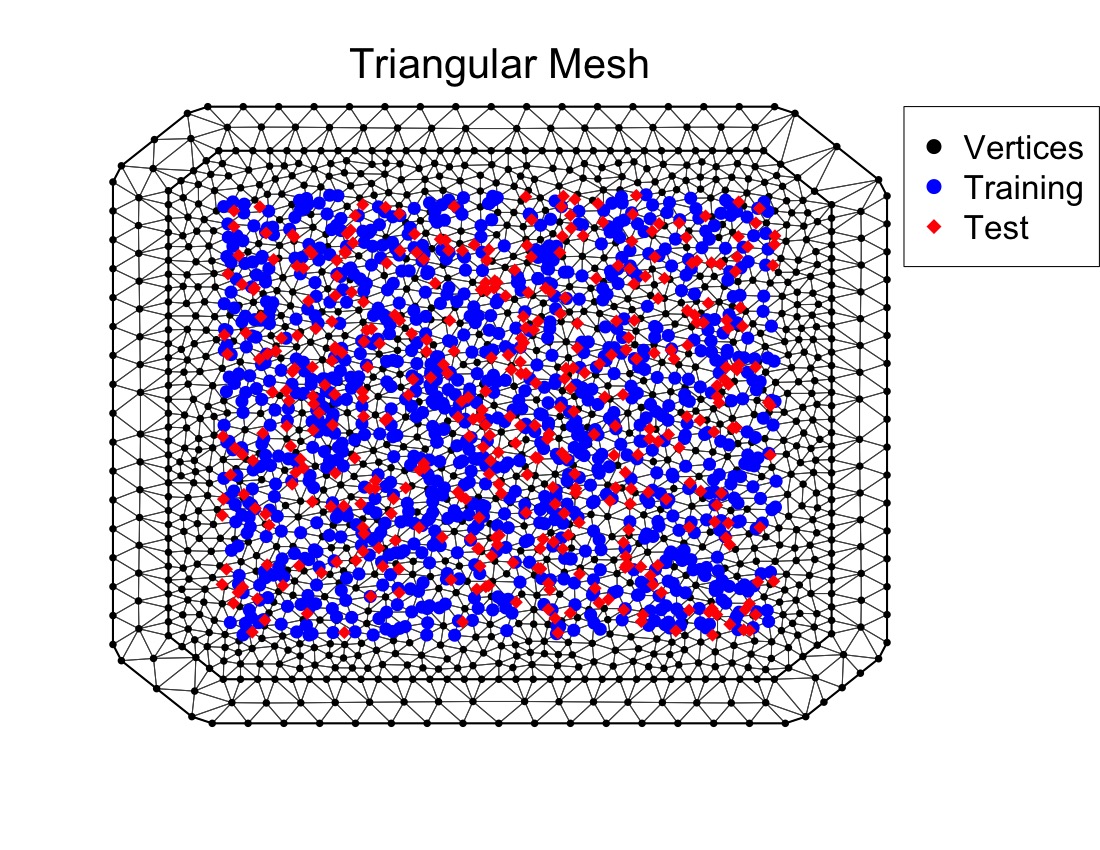}
    \caption{Triangular Mesh for data in simulation studies. Black points denote the vertices, or nodes, of the triangular mesh. Blue points represent the observation locations used to fit the hierarchical spatial models, and the red points denote the observations locations for the validation sample. }
    \label{fig:mesh}
\end{figure}

\begin{figure}[ht]
    \centering
    \includegraphics[width=0.7\textwidth]{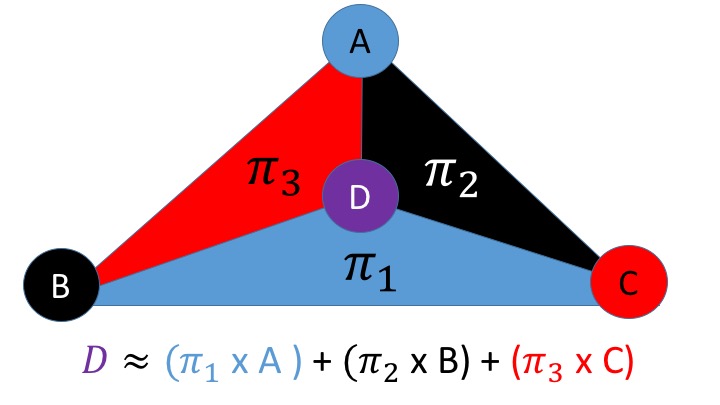}
        \caption{Diagram of the piece-wise linear basis functions. Point $D$ is the observation location, points $A$, $B$, and $C$ are the triangle vertices, and $\pi_{1},\pi_{2},$ and $\pi_{3}$ are the corresponding weights.  The weights $\pi_{1}$, $\pi_{2}$, and $\pi_{3}$ correspond to the proportion of the area of the specified triangle to the area of the larger triangle. We interpolate point $D$ by taking the weighted mean of the three triangle vertices where $D\approx \pi_{1}A + \pi_{2}B + \pi_{3}C$}
        \label{Fig:PieceWise}
\end{figure}

\begin{table}[ht]
\centering
\begin{tabular}{lllrr}
  \hline
  Precision  \\ 
Matrix & $\beta_{1}$ (95\% CI) & $\beta_{2}$ (95\% CI) & CVMPSE & Time (min) \\ 
  \hline
Ind & 1.07 (0.8,1.34) & 0.92 (0.65,1.18) & 0.28 & 9.53 \\ 
  ICAR & 1.09 (0.82,1.37) & 0.93 (0.67,1.2) & 0.27 & 10.73 \\ 
  CAR & 1.05 (0.79,1.33) & 0.91 (0.65,1.18) & 0.27 & 10.38 \\ 
Gold Standard & 1.03 (0.77,1.3) & 0.89 (0.63,1.16) & 0.29 & 3624.43 \\ 
   \hline
\end{tabular}
\caption{Simulated example with binary spatial observations. Parameter estimation, prediction, and model fitting time results across precision matrices.} 
\label{tab:binprec}
\end{table}

\begin{figure}
    \centering
    \includegraphics[width=0.9\textwidth]{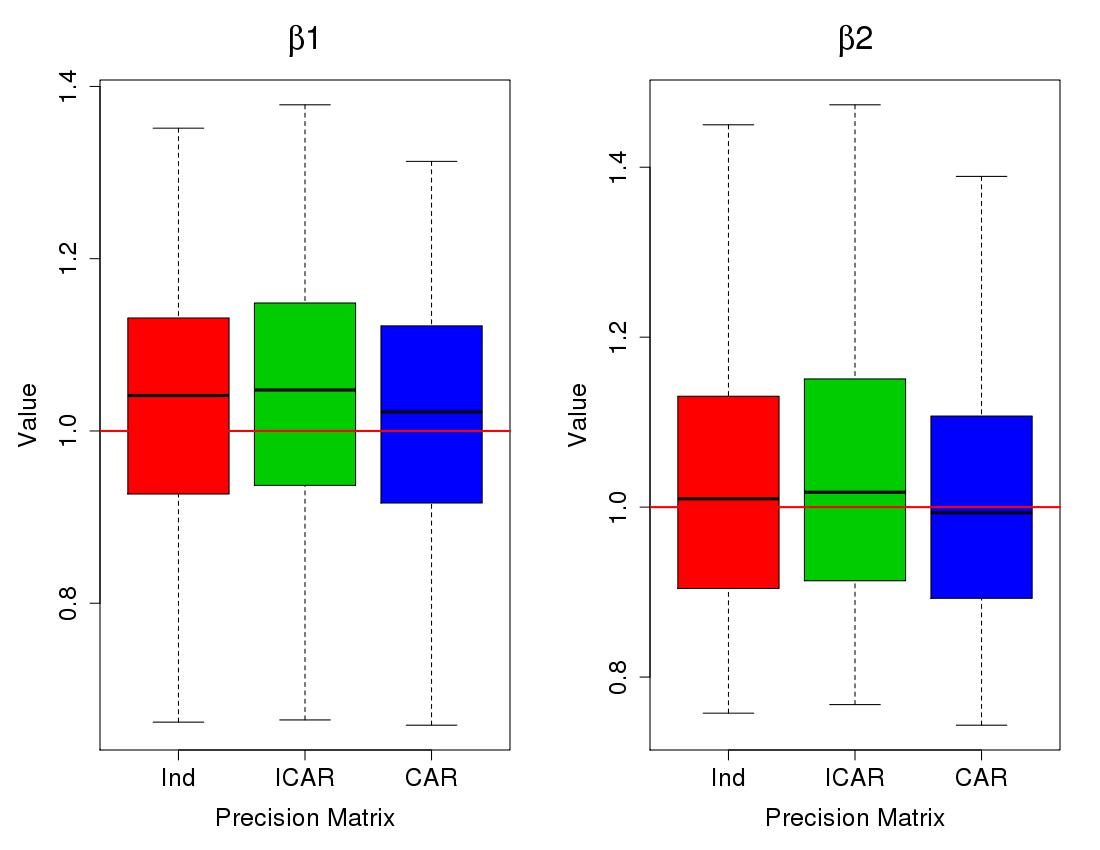}
    \caption{Binary data simulation study: distribution of posterior mean estimates for parameters $\beta_{1}$ (left) and $\beta_{2}$ (right) for three different precision matrices - Independent (red), ICAR (green), and CAR with $\phi=0.5$ (blue). The suitable rank $p$ of the Moran's operator $\mathbf{M}$ chosen using the automated heuristic. Distributions are similar across precision matrices.}
    \label{Fig:BinaryBoxplots}
\end{figure}

\begin{table}[ht]
\centering
\begin{tabular}{rrr}
  \hline
 & $\beta_{1}$ & $\beta_{2}$ \\ 
  \hline
Independent & 0.88 & 0.93 \\ 
  ICAR & 0.89 & 0.91 \\ 
  CAR & 0.88 & 0.95 \\ 
   \hline
\end{tabular}
\caption{Binary data simulation study: Coverage probabilities for 100 simulated samples. 
Columns correspond to the regression coefficients. Rows correspond to the type of precision matrix.} 
\label{Table:BinaryCoverage}
\end{table}

\section{Simulation study with spatial count observations}

We conduct a simulation study using spatial count observations using 100 samples. The regression coefficients and the latent spatial random field are generated in the same way as the binary case. The observations come from a spatial generalized linear mixed model (SGLMM) with a Poisson data model and a log link function. Mesh construction and model fitting details follow closely to the binary case. We select one sample (from the 100 generated samples) as the dataset for the comparative analysis. When comparing across ranks, we elect to use the precision matrix from the intrinsic conditional auto-regressive model (ICAR) model $\mathbf{Q}=(\mbox{diag}(\mathbf{N1})-\mathbf{N})$, where $\mathbf{N}$ is the adjacency or neighborhood matrix and $\mathbf{1}$ is $m$-dimensional vector of 1s. We also compare inferential and predictive performance across the three different precision matrices as in the binary case. 

Results indicate that the choice of rank (for the Moran's operator) is a key driver for accurate parameter estimation and prediction as noted in Table \ref{Table:PoissonRank}. As in the binary case, the choice of precision matrices does not influence inference or prediction as shown in Table \ref{Table:PoissonPrec}. Coverage probabilities (Table \ref{Table:PoissonCoverage}) align with the nominal coverage ($95\%$).  The PICAR approach improves mixing in the MCMC algorithm as shown by the larger effective samples per second (ESS/sec) compared to the gold standard approach. For model parameters $\beta_{1}$ and $\beta_{2}$, PICAR yields an ESS/sec of $6.4$ and $7.2$ respectively and the gold standard returns an ESS/sec $0.09$ and $0.09$ respectively. For the random effects $\bs{W}$, the average ESS/sec is $1.8$ for the PICAR approach and $0.018$ for the gold standard, an improvement by a factor of roughly $101$.

\begin{table}[ht]
\centering
\begin{tabular}{lllrr}
  \hline
Rank & $\beta_{1}$ (95\% CI) & $\beta_{2}$ (95\% CI)& CVMPSE & Time (min) \\ 
  \hline
  10 & 1.09 (0.99,1.19) & 1.01 (0.92,1.11) & 1.96 & 8.84 \\ 
  50 & 1.05 (0.95,1.15) & 1.02 (0.92,1.12) & 1.74 & 9.87 \\ 
  \textbf{62} & \textbf{1.04 (0.94,1.14)} & \textbf{0.99 (0.89,1.09)} & \textbf{1.57} & \textbf{10.65} \\ 
  75 & 1.03 (0.93,1.14) & 0.99 (0.89,1.09) & 1.66 & 10.39 \\ 
  100 & 1.05 (0.95,1.16) & 0.98 (0.88,1.09) & 1.71 & 11.07 \\ 
  200 & 1.08 (0.97,1.19) & 0.98 (0.87,1.1) & 1.81 & 13.49 \\ 
  Gold Standard & 1.07 (0.97,1.17) & 1.01 (0.91,1.12) & 1.66 & 3803.84 \\ 
  
   \hline
\end{tabular}
\caption{Simulated example with count spatial observations. Parameter estimation, prediction, and model fitting time results across Moran's basis ranks. Bold font denotes the rank chosen by the automated heuristic. } 
\label{Table:PoissonRank}
\end{table}

\begin{table}[ht]
\centering
\begin{tabular}{lllrr}
  \hline
  Precision  \\ 
Matrix & $\beta_{1}$ (95\% CI) & $\beta_{2}$ (95\% CI) & CVMPSE & Time (min) \\ 
  \hline
Ind & 1.04 (0.94,1.14) & 0.99 (0.89,1.09) & 1.56 & 10.92 \\ 
  ICAR & 1.04 (0.94,1.14) & 0.99 (0.89,1.09) & 1.57 & 10.65 \\ 
  CAR & 1.04 (0.94,1.15) & 0.99 (0.89,1.09) & 1.57 & 10.17 \\ 
 Gold Standard & 1.07 (0.97,1.17) & 1.01 (0.91,1.12) & 1.66 & 3803.84 \\ 
   \hline
\end{tabular}
\caption{Simulated example with count spatial observations. Parameter estimation, prediction, and model fitting time results across precision matrices.} 
\label{Table:PoissonPrec}
\end{table}

\begin{figure}
    \centering
    \includegraphics[width=0.9\textwidth]{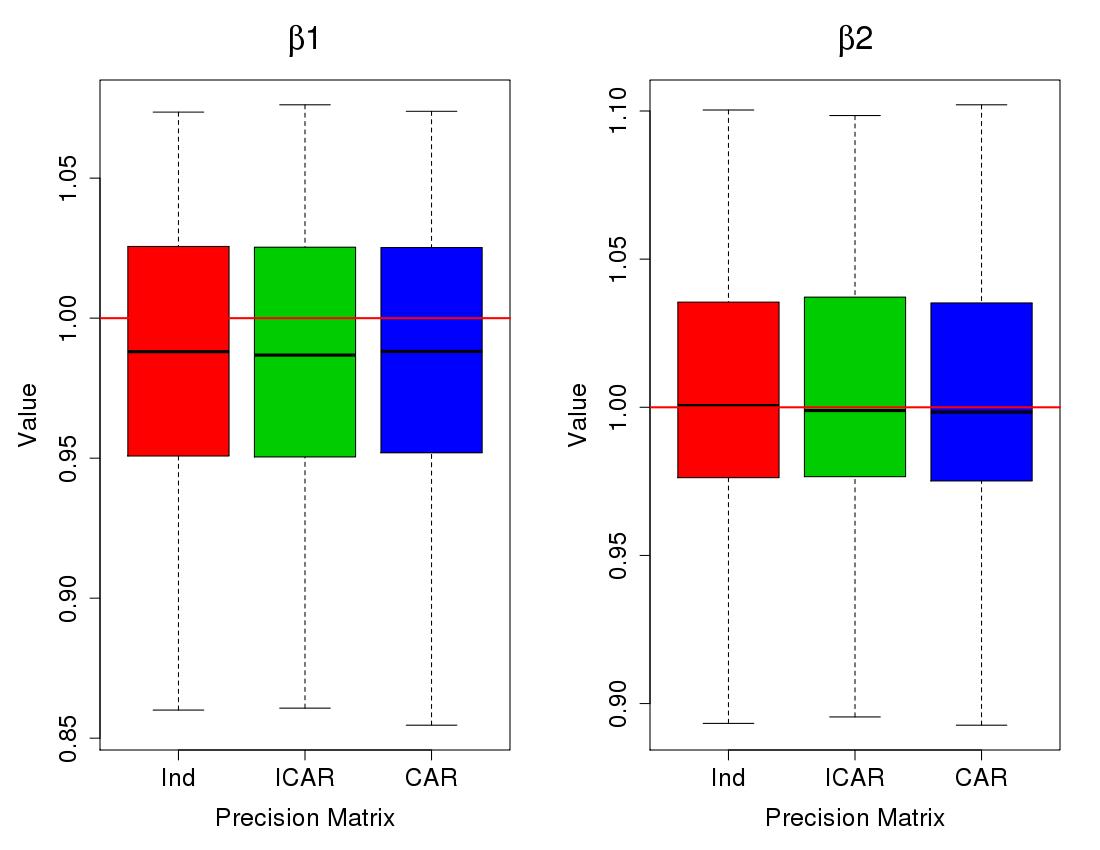}
    \caption{Poisson data simulation study: distribution of posterior mean estimates for parameters $\beta_{1}$ (left) and $\beta_{2}$ (right) for three different precision matrices - Independent (red), ICAR (green), and CAR with $\phi=0.5$ (blue). The suitable rank $p$ of the Moran's operator $\mathbf{M}$ chosen using the automated heuristic. Distributions are similar across precision matrices.}
    \label{Fig:PoissonBoxplots}
\end{figure}

\begin{table}[ht]
\centering
\begin{tabular}{rrr}
  \hline
 & $\beta_{1}$ & $\beta_{2}$ \\ 
  \hline
Independent & 0.95 & 0.97 \\ 
  ICAR & 0.95 & 0.96 \\ 
  CAR & 0.95 & 0.95 \\ 
   \hline
\end{tabular}
\caption{Poisson data simulation study: Coverage probabilities for 100 simulated samples. 
Columns correspond to the regression coefficients. Rows correspond to the type of precision matrix.} 
\label{Table:PoissonCoverage}
\end{table}

\begin{landscape}
\begin{table}[ht]
\centering
\begin{tabular}{rllllll}
  \hline
 & 1 & 2 & 3 & 4 & 5 & 6 \\ 
 Rank & $\beta_{1}$ (95\% CI) & $\beta_{2}$ (95\% CI)& $\alpha_{1}$ (95\% CI)& $\alpha_{2}$ (95\% CI)& Mismatch \% & Time (min) \\ 
  \hline
10 & 0.88 (0.64,1.11) & 1.01 (0.77,1.24) & -0.03 (-0.17,0.11) & -0.02 (-0.2,0.15) & 0.43 & 12.77 \\ 
  \textbf{23} & \textbf{0.9 (0.67,1.15)} & \textbf{1.05 (0.81,1.28)} & \textbf{0 (-0.14,0.14)} & \textbf{0 (-0.17,0.17)} & \textbf{0.42} & \textbf{12.64} \\ 
  50 & 0.94 (0.69,1.18) & 1.12 (0.87,1.36) & 0.05 (-0.09,0.18) & 0.05 (-0.13,0.22) & 0.41 & 13.64 \\ 
  75 & 0.98 (0.73,1.23) & 1.17 (0.92,1.43) & 0.07 (-0.07,0.21) & 0.07 (-0.1,0.25) & 0.41 & 14.63 \\ 
  100 & 1.01 (0.77,1.27) & 1.22 (0.96,1.48) & 0.09 (-0.05,0.22) & 0.09 (-0.09,0.26) & 0.4 & 16.7 \\ 
  200 & 1.12 (0.85,1.4) & 1.24 (0.97,1.52) & 0.15 (0.01,0.29) & 0.13 (-0.05,0.3) & 0.41 & 18.92 \\ 
  300 & 1.23 (0.94,1.52) & 1.35 (1.06,1.64) & 0.22 (0.08,0.35) & 0.19 (0.02,0.36) & 0.41 & 20.94 \\ 
  Gold Standard & 0.9 (0.65,1.12) & 1.07 (0.82,1.31) & 0.02 (-0.12,0.16) & 0.02 (-0.17,0.18) & 0.42 & 7558.88 \\ 
   \hline
\end{tabular}
\caption{Simulated example with ordered categorical spatial observations. Parameter estimation, prediction, and model fitting time results across Moran's basis ranks. Bold font denotes the rank chosen by the automated heuristic.}
\label{Table:OrdinalRank}
\end{table}

\end{landscape}

\begin{landscape}
\begin{table}[ht]
\centering
\begin{tabular}{rllllll}
  \hline
 & 1 & 2 & 3 & 4 & 5 & 6 \\ 
 Rank & $\beta_{1}$ (95\% CI) & $\beta_{2}$ (95\% CI)& $\alpha_{1}$ (95\% CI)& $\alpha_{2}$ (95\% CI)& Mismatch \% & Time (min) \\ 
  \hline
Ind & 0.9 (0.66,1.14) & 1.04 (0.8,1.28) & 0.01 (-0.13,0.14) & 0 (-0.17,0.17) & 0.42 & 11.70 \\ 
  ICAR & 0.9 (0.67,1.15) & 1.05 (0.81,1.28) & 0 (-0.14,0.14) & 0 (-0.17,0.17) & 0.42 & 12.63 \\ 
  CAR & 0.89 (0.65,1.12) & 1.03 (0.79,1.26) & 0.01 (-0.13,0.15) & 0 (-0.17,0.18) & 0.43 & 12.87 \\
 Gold Standard & 0.9 (0.65,1.12) & 1.07 (0.82,1.31) & 0.02 (-0.12,0.16) & 0.02 (-0.17,0.18) & 0.42 & 7558.88 \\ 
     \hline
\end{tabular}
\caption{Simulated example with ordered categorical spatial observations. Parameter estimation, prediction, and model fitting time results across precision matrices.}
\label{Table:OrdinalPrec}
\end{table}
\end{landscape}

\begin{figure}[ht]
    \centering
    \includegraphics[width=0.9\textwidth]{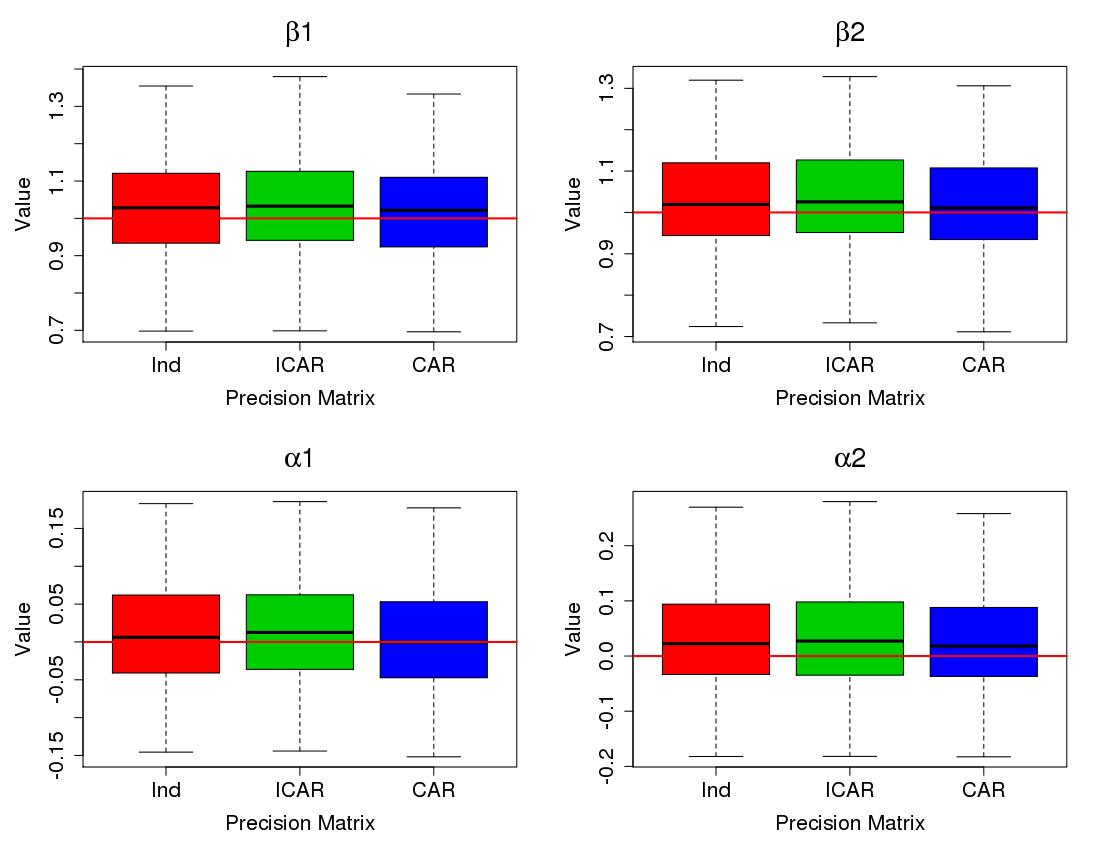}
    \caption{Ordinal data simulation study: distribution of posterior mean estimates for parameters $\beta_{1}$ (top left) $\beta_{2}$ (top right), $\alpha_{1}$ (bottom left), and $\alpha_{2}$ (bottom right) for three different precision matrices - Independent (red), ICAR (green), and CAR with $\phi=0.5$ (blue). The red horizontal line denotes the true parameter values. The automated heuristic selects the appropriate rank $p$ of the Moran's operator $\mathbf{M}$. Note that the default precision matrix for the PICAR approach is the ICAR precision matrix (green). Distributions are similar across precision matrices.}
    \label{Fig:OrdinalBoxplots}
\end{figure}

\begin{table}[ht]
\centering
\begin{tabular}{rrrrr}
  \hline
 & $\beta_{1}$ & $\beta_{2}$ & $\alpha_{1}$ & $\alpha_{2}$\\ 
  \hline
Independent & 0.92 & 0.94 & 0.93 & 0.90 \\ 
  ICAR & 0.91 & 0.92 & 0.93 & 0.88 \\ 
  CAR & 0.93 & 0.96 & 0.93 & 0.94 \\ 
   \hline
\end{tabular}
\caption{Ordered categorical data simulation study: Coverage probabilities for 100 simulated samples. 
Columns correspond to the regression coefficients. Rows correspond to the type of precision matrix.} 
\label{Table:OrdinalCoverage}
\end{table}

\begin{table}[ht]
\centering
\begin{tabular}{lllrr}
  \hline
Rank & $\beta_{1}$ (95\% CI) & $\beta_{2}$ (95\% CI)& CVMPSE & Time (min) \\ 
  \hline
10 & 0.92 (0.81,1.02) & 0.94 (0.85,1.03) & 3.59 & 0.52 \\ 
  50 & 0.77 (0.54,1.01) & 0.99 (0.89,1.09) & 2.30 & 7.68 \\ 
  \textbf{63} & \textbf{0.77 (0.48,1.06)} & \textbf{1.02 (0.9,1.12)} & \textbf{2.31} & \textbf{13.15} \\ 
  75 & 0.86 (0.55,1.16) & 1.03 (0.93,1.14) & 2.68 & 22.07 \\ 
  100 & 0.95 (0.55,1.33) & 1.06 (0.94,1.17) & 2.82 & 46.55 \\ 
  200 & 1.07 (0.7,1.49) & 1.07 (0.95,1.19) & 3.80 & 226.49 \\ 
   \hline
\end{tabular}

\caption{Simulated example with spatially varying coefficients. Model fit using {\tt stan} programming language. Parameter estimation, prediction, and model fitting time results across Moran's basis ranks. Bold font denotes the rank chosen by the automated heuristic. } 
\label{Table:SVCRank}
\end{table}

\section{Comparative Study of Basis Functions for the Spatially Varying Coefficients Model}
Here, we provide additional details regarding the basis functions from the comparative study (Section 4.2). We examine four classes of basis functions, the Moran's basis functions, eigenvectors of a pre-specified Mat\'ern covariance function, bi-square basis functions, and thin-plate splines. The Moran's basis functions are generated from a mesh comprised of $m=1,137$ vertices. Using the automated heuristic (Section 3.3), the Moran's basis function uses the $p_{\mbox{Mor}}=105$ leading eigenvectors of the Moran's operator. Using the same mesh, we generate a new set of basis functions based on a pre-specified Mat\'ern covariance function. The finite set of locations are the $m$-dimensional mesh vertices, and the covariance parameters are range $\phi=0.2$, partial sill $\sigma^{2}=1$, and smoothness $\nu=0.5$. The automated heuristic chose the $p_{\mbox{Mat}}=102$ leading eigenvectors of the covariance function.

Thin-plate splines and the bi-square basis functions are constructed using a grid of $64$ evenly-spaced knot locations in the spatial domain. We chose $64$ knot locations, as opposed to $25, 36, 49, 81, 100,$ or $121$ based on out-of-sample prediction error. The bi-square basis functions are defined as: 
$$\pmb{\Phi}_{BS}(\bs{s})=\Bigg\{1-\Bigg(\frac{||\bs{s}-\bs{c}||}{\omega}\Bigg)^{2}\Bigg\}^{2} I(||\bs{s}-\bs{c}||<\omega),$$
where $\bs{s}$ is the observation location and $\bs{c}$ is knot location or the center of the basis function. The bi-square basis functions have locally bounded support $\{\bs{s} \in \mathbb{R}^{2}: ||\bs{s}-\bs{c}||<\omega\}$ that is controlled by the radius, or threshold, $\omega>0$. In the comparative study, we fix $\omega=0.3$. The thin plate splines, a special class of polyharmonic splines, are defined as:
$$\pmb{\Phi}_{TPS}(\bs{s})=||\bs{s}-\bs{c}||^{2}\log||\bs{s}-\bs{c}||.$$
 $\bs{s}$ is the observation location and $\bs{c}$ is knot location or the center of the basis function. Note that thin-plate splines do not have a tuning parameter such as $\omega$ in the bi-square basis functions.

\begin{table}[ht]
\centering
\begin{tabular}{rllrr}
  \hline
 & $\beta_{1}$ & $\beta_{2}$ & CVMSPE & Basis \\ 
  \hline
PICAR & 1.13 (1.05,1.2) & 1.03 (0.97,1.09) & 3.43 & 102 \\ 
  Eigenvector & 1.21 (1.13,1.28) & 1.04 (0.99,1.1) & 3.95 & 105 \\ 
  Bisquare & 0.99 (-0.01,2.14) & 0.96 (0.9,1.01) & 3.60 &  64 \\ 
  Thin Plate Splines & 0.34 (-0.79,1.59) & 0.94 (0.88,0.99) & 4.71 &  64 \\ 
   \hline
\end{tabular}
   \caption{Comparative Study among basis-representation approaches, PICAR, eigenvectors of a Mat\'ern covariance function, bi-square basis functions, and thin-plate splines.}
\end{table}
\medskip

\section{Simulation Study: Mesh Densities and Smoothness}
In the following numerical study, we examine how PICAR performs under varying mesh densities and smoothness of the latent spatial random field. We generate $n=1000$ observations for model-fitting and $n_{cv}=300$ observations for validation. We fix regression coefficients $\beta=(1,1)^\prime$ and Mat\'ern covariance function parameters $\sigma^{2}=1$ and $\phi=0.2$ with varying smoothness $\nu$. Additional simulation study details are as follows:
    \begin{enumerate}
        \item Data Types: Binary and Count
        \item Smoothness of the Mat\'ern Covariance Function (3): $\nu\in\{0.5, 2.5, \infty\}$
        \item Mesh Density (6): $m=\{100,500,750,1000,1500,2000\}$
    \end{enumerate}
The latent spatial random field is a zero-mean Gaussian process with covariance function from the Mat\'ern class. For each data type (count and binary), we generate three different datasets corresponding to a value of $\nu$, specifically $\nu=\{0.5,2.5,\infty\}$. The case where $\nu=0.5$ returns a `rougher' exponential covariance function, $\nu=\infty$ yields the infinitely differentiable smooth Gaussian covariance functions, and $\nu=2.5$ corresponds to mid-range `smoothness.' 
The data are generated according to Section 4 and Supplement Section 2 for binary and count data, respectively. In total, we generate six observational datasets and fit six different PICAR-based models with varying mesh dimensions to each dataset. We model the observations using PICAR and use the automated heuristic (Section 3.3) to select a suitable rank $p$. 

For each case, we compare the predictive accuracy (mean squared prediction error) and the associated prediction standard deviations (average standard deviations of our predictions). Out-of-sample prediction results are presented in Table \ref{Table:MeshBinary} and Table \ref{Table:MeshCount} for the binary and count data, respectively. We provide the prediction standard deviation results in Table \ref{Table:MeshBinarySD} and Table \ref{Table:MeshCountSD}, respectively, for the binary and count data. For count data, a denser mesh typically leads to more accurate predictions, but may inflate the prediction standard deviations. For the binary case, we observe comparable prediction errors and predictive standard deviations across varying mesh densities.

We also examine the chosen rank for each dataset. For count observations, we find that the chosen rank $p$ decreases as the latent spatial random field becomes smoother (i.e. as $\nu$ increases). For smoother random fields, fewer basis functions are needed to represent the underlying spatial processes. For the binary case, PICAR selects a lower rank $p$ when $\nu=0.5$ and $\nu=\infty$ and a higher rank when $\nu=2.5$. This pattern holds across varying mesh densities $m$. 
 
These results suggest that predictive ability (prediction error and precision) and rank selection is sensitive to the type of spatial data (e.g. count or binary), the smoothness of the underlying spatial random field ($\nu$), and the mesh density ($p$). Given these nuances, a more extensive study would be needed to thoroughly examine the effect of mesh density on predictions. While this is outside the scope of this manuscript, a closer examination of mesh design and rank selection would be a fruitful avenue for future research. In practice, we suggest practitioners to model their datasets with varying mesh densities and then choose the appropriate mesh density ($m$) and rank ($p$) based on the lowest out-of-sample prediction error.

\begin{table}[ht]
\centering
\begin{tabular}{rrrrrrr}
  \hline
  & \multicolumn{6}{c}{Mesh density $m$} \\ 
$\nu$ & 100 & 500 & 750 & 1000 & 1500 & 2000 \\ 
  \hline
0.5 & 0.38 & 0.36 & 0.35 & 0.38 & 0.34 & 0.36 \\ 
  2.5 & 0.33 & 0.32 & 0.33 & 0.32 & 0.32 & 0.36 \\ 
  $\infty$ & 0.32 & 0.32 & 0.32 & 0.31 & 0.32 & 0.33 \\ 
   \hline
\end{tabular}
\caption{Mesh design sensitivity analysis for binary observations. Table contains the average misclassification error for the validation sample. Columns represent the mesh densities ranging from 100 to 2,000. Rows denote the smoothness of the latent spatial random field.} 
\label{Table:MeshBinary}
\end{table}

\begin{table}[ht]
\centering
\begin{tabular}{rrrrrrr}
  \hline
   & \multicolumn{6}{c}{Mesh density $m$} \\ 
 $\nu$ & 100 & 500 & 750 & 1000 & 1500 & 2000 \\ 
  \hline
0.5 & 11.45 & 8.04 & 7.54 & 8.95 & 9.04 & 7.80 \\ 
  2.5 & 6.40 & 4.34 & 4.23 & 4.87 & 4.42 & 5.85 \\ 
  $\infty$ & 2.58 & 2.54 & 2.38 & 2.31 & 2.13 & 2.46 \\ 
   \hline
\end{tabular}
\caption{Mesh design sensitivity analysis for count observations. Table contains the average mean squared prediction error for the validation sample. Columns represent the mesh densities ranging from 100 to 2,000. Rows denote the smoothness of the latent spatial random field.} 
\label{Table:MeshCount}
\end{table}

\begin{table}[ht]
\centering
\begin{tabular}{rrrrrrr}
  \hline
  & \multicolumn{6}{c}{Mesh density $m$} \\ 
$\nu$ & 100 & 500 & 750 & 1000 & 1500 & 2000 \\ 
  \hline
0.5 & 0.07 & 0.09 & 0.07 & 0.10 & 0.09 & 0.07 \\ 
  2.5 & 0.07 & 0.09 & 0.09 & 0.09 & 0.09 & 0.08 \\ 
  $\infty$ & 0.04 & 0.06 & 0.06 & 0.07 & 0.06 & 0.04 \\ 
   \hline
\end{tabular}
\caption{Mesh design sensitivity analysis for binary observations. Table contains the average prediction standard deviation for the validation sample. Columns represent the mesh densities ranging from 100 to 2,000. Rows denote the smoothness of the latent spatial random field.} 
\label{Table:MeshBinarySD}
\end{table}

\begin{table}[ht]
\centering
\begin{tabular}{rrrrrrr}
  \hline
   & \multicolumn{6}{c}{Mesh density $m$} \\ 
 $\nu$ & 100 & 500 & 750 & 1000 & 1500 & 2000 \\ 
  \hline
0.5 & 0.31 & 0.42 & 0.45 & 0.42 & 0.45 & 0.45 \\ 
  2.5 & 0.30 & 0.42 & 0.39 & 0.43 & 0.42 & 0.45 \\ 
  $\infty$ & 0.24 & 0.34 & 0.30 & 0.33 & 0.33 & 0.34 \\ 
   \hline
\end{tabular}
\caption{Mesh design sensitivity analysis for count observations. Table contains the average prediction standard deviation for the validation sample. Columns represent the mesh densities ranging from 100 to 2,000. Rows denote the smoothness of the latent spatial random field.} 
\label{Table:MeshCountSD}
\end{table}

\begin{table}[ht]
\centering
\begin{tabular}{rrrrrrr}
  \hline
 $\nu$ & 100 & 500 & 750 & 1000 & 1500 & 2000 \\ 
  \hline
0.5 &  27 &  57 &  20 &  93 &  50 &  20 \\ 
  2.5 &  43 &  97 &  96 &  89 &  94 &  49 \\ 
  $\infty$ &   7 &  23 &  18 &  23 &  18 &   5 \\ 
   \hline
\end{tabular}
\caption{Simulation study results for the chosen rank $p$ for the binary observations. Rows denote the smoothness ($\nu$) of the latent spatial random field, and columns represent the mesh density, or the number of mesh nodes.} 
\label{Table:RankBinary}
\end{table}

    \begin{table}[ht]
\centering
\begin{tabular}{rrrrrrr}
  \hline
 $\nu$ & 100 & 500 & 750 & 1000 & 1500 & 2000 \\ 
  \hline
0.5 &  45 &  98 & 101 &  97 &  99 &  94 \\ 
  2.5 &  45 &  82 &  76 &  95 &  95 &  98 \\ 
  $\infty$ &  50 &  83 &  67 &  78 &  85 &  85 \\ 
   \hline
\end{tabular}
\caption{Simulation study results for the chosen rank $p$ for the count observations. Rows denote the smoothness ($\nu$) of the latent spatial random field, and columns represent the mesh density, or the number of mesh nodes.} 
\label{Table:RankCount}
\end{table}

\section{Binary Data: Parasitic Infestation of Dwarf Mistletoe}
In Minnesota, the eastern spruce dwarf mistletoe (\textit{Arceuthobium pusillum}) are a parasitic species that affect the longevity and quality of its host, the black spruce (Picea mariana) \citep{geils2002damage}. This infestation has economic ramifications because the black spruce are valuable resources used to produce high quality paper. We use data from the Minnesota Department of Natural Resources (DNR) forest inventory \citep[cf.][]{hanks2011}. The response is a binary incidence of dwarf mistletoe at $n=25,431$ black spruce stands (location hosting the black spruce samples). We randomly sample $22,888$ observations to fit our model and reserve $2,543$ observations for validation. Covariates include: (1) average age of trees in the stand; (2) basal area per acre of trees in the stand; (3) average canopy height; and (4) volume of the stand in cords, a unit of measurement. We fit the binary spatial model of Section 4.1 and construct a triangular mesh with $m=32,611$ vertices, using our heuristic (Section 3.3) to select a rank of $p=520$ for the Moran's basis functions matrix. 

\begin{figure}[ht]
    \centering
    \includegraphics[width=0.9\textwidth]{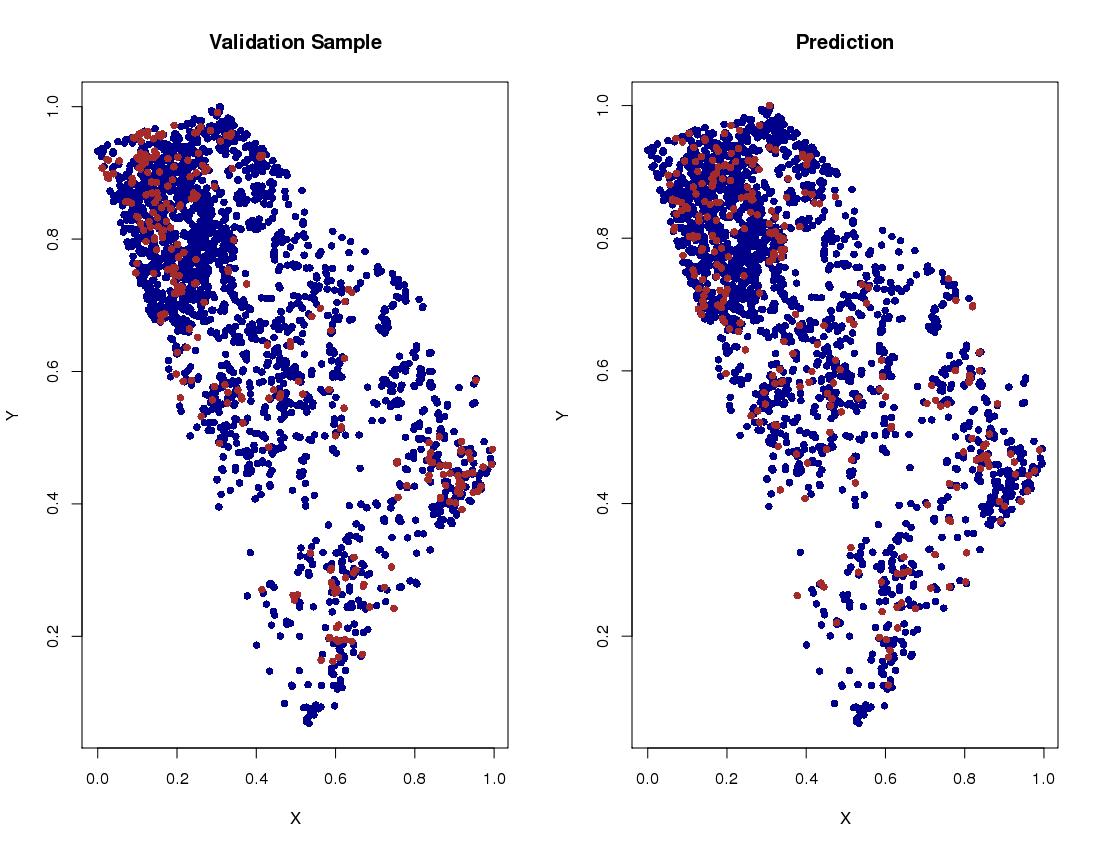}
    \caption{Observed (left) and predicted (right) dwarf mistletoe presence and absence at the validation sample locations. Red points denote the presence of dwarf mistletoe and blue points denote absence.}
    \label{Fig:MistletoePrediction}
\end{figure}

The PICAR approach required around $4$ hours: $2$ hours to run $10^{5}$ iterations of the MCMC algorithm, $10$ minutes to generate the Moran's operator (Section 3.1) via parallel computing across 100 processors, and $1.7$ hours to calculate the first $1,000$ eigencomponents using the \texttt{RSpectra} package \citep{RSpectra2019}. Comparison with the full SGLMM is computationally infeasible. The posterior predictive map displays similar spatial patterns between the predicted values and the validation sample.
(Figure \ref{Fig:MistletoePrediction}). 

\begin{table}[ht]
\centering
\begin{tabular}{rrl}
  \hline
 Covariate & Estimate & 95\% CI \\ 
  \hline
  Age & 0.0008 & (-0.0041,0.0034) \\ 
  Basal Area & -0.0045 & (-0.007,-0.0026) \\ 
  Height & 0.0203 & (0.0157,0.0236) \\ 
  Volume & -0.0026 & (-0.0034,-0.0017) \\ 
  tau & 0.0040 & (0.0021,0.0094) \\ 
  
   \hline
\end{tabular}
\caption{Inference results for the mistletoe data. Rows correspond to the predictor variables and columns include the parameter estimates and 95\% credible intervals}
\label{Table:Mistletoe}
\end{table}

\begin{figure}
    \centering
    \includegraphics[width=0.9\textwidth]{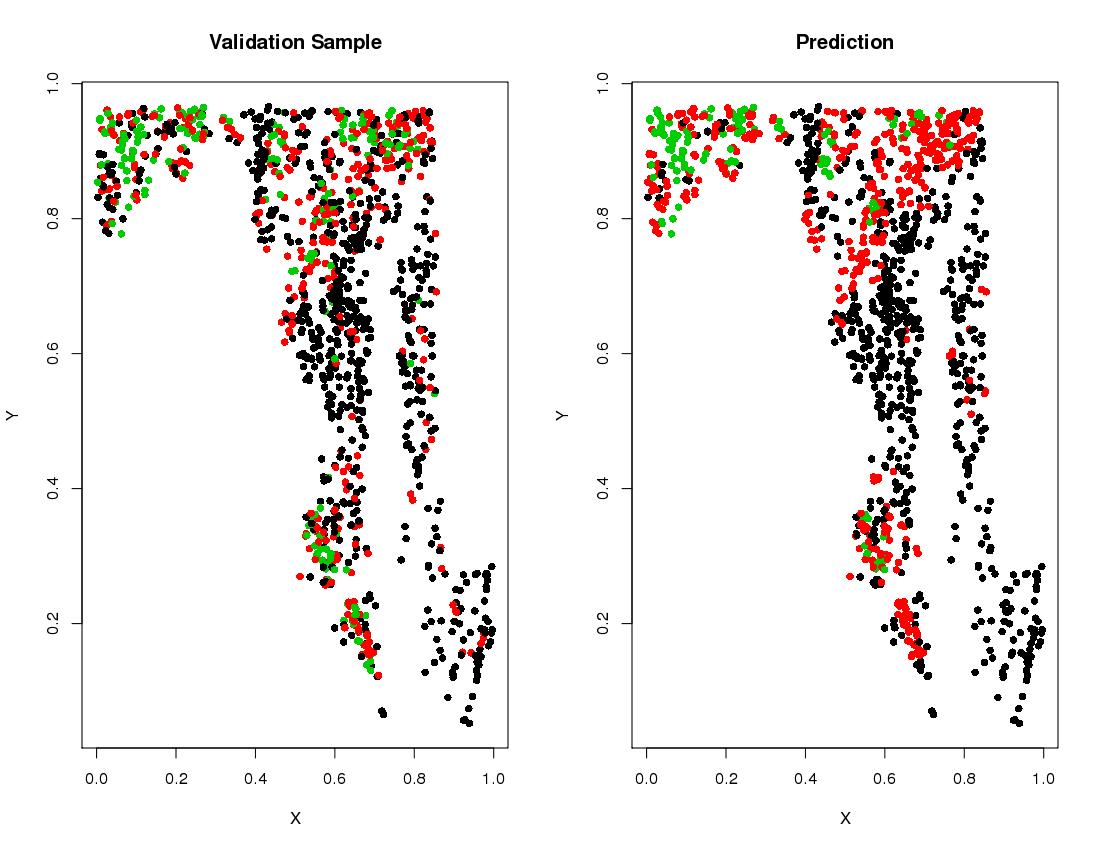}
    \caption{The left panel shows the BIBI index at the prediction locations and the right panel shows the predicted BIBI index. Black, red, and green points indicate low, medium and high levels of BIBI respectively.}
    \label{Fig:MDWadersPrediction}
\end{figure}

\section{Sparse Matrix Computations}
We use the sparse matrix R package \textbf{Matrix} \citep{MatrixPackage2019} to reduce costs for the operation $\pmb{\Sigma}=\mathbf{(I-11'}/m)\mathbf{N}$, where $\bs{N}$ is the neighborhood matrix of the mesh vertices. Then, we partition the resulting matrix $\pmb{\Sigma}$ into $K$ mutually exclusive $\frac{1}{K}\times n$ sub-matrices $\pmb{\Sigma}_{k}$. By parallelizing across $K$ processors, we can quickly construct the partial Moran's operator $\bs{MO}_{k}=\pmb{\Sigma}_{k}(I-11'/m)$ for $k=1,...,K$. Finally, we generate the full Moran's operator by combining the $\bs{MO}_{k}$'s as so $\bs{MO}=\begin{bmatrix} 
\bs{MO}_{1}\\
\vdots\\
\bs{MO}_{K}
\end{bmatrix}$.
We can compute the leading $k$ eigencomponents of the Moran's operator using a partial eigendecomposition approach such as the Implicitly Restarted Arnoldi Method \citep{lehoucq1998arpack} from \texttt{RSpectra} package \citep{RSpectra2019}.

\begin{figure}[ht]
    \centering
    \includegraphics[width=0.9\textwidth]{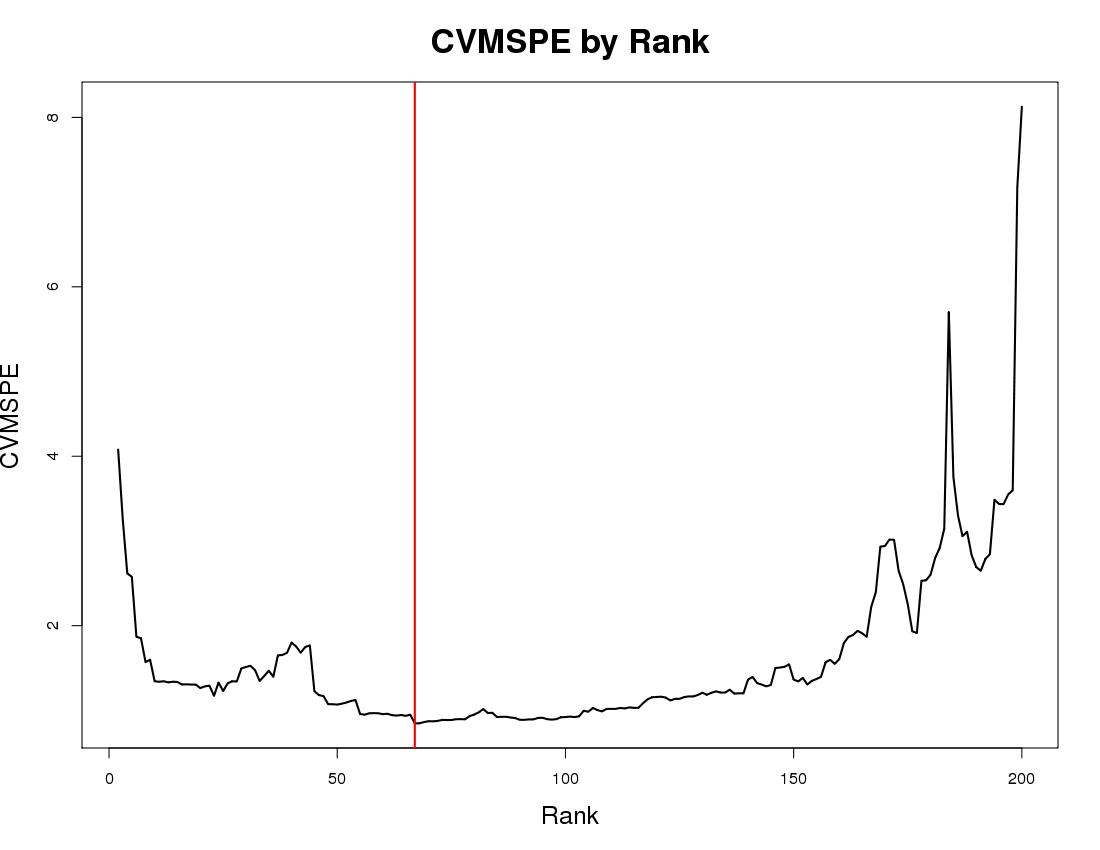}
    \caption{Cross-validated mean squared prediction error (CVMSPE) for ranks 1-200 using the automated heuristic. The vertical red line denotes the chosen rank ($p=63$) with lowest CVMPSE.}
    \label{Fig:RankSelection}
\end{figure}

\bibliography{references}